\documentclass[usenatbib]{mn2e}
\usepackage{myaasmacros}
\usepackage{graphicx}
\usepackage{ulem}
\usepackage{amsmath}
\usepackage{amssymb}
\usepackage{bm}
\usepackage[table,dvipsnames]{xcolor}

\definecolor{darkgreen}{rgb}{0.0,0.5,0.0}
\definecolor{darkred}{rgb}{0.5,0.0,0.0}
\definecolor{brown}{rgb}{0.65,.16,0.16}
\definecolor{grey}{rgb}{0.4,0.5,0.6}

\newcommand{\twocol}[1]{\multicolumn{2}{c}{\makebox[1.5cm][l]{#1}}}

\def\siglos{\sigma_{\text{LOS}}}

\def\dd{\text{d}}
\def\betastar{\widetilde{\beta}}
\def\DM{\text{D}}
\def\vlosfour{\langle v_{\rm LOS}^4 \rangle}

\def\vsone{v_{s1}}
\def\vstwo{v_{s2}}

\def\GravSphere{{\sc GravSphere}}
\def\DiscreteJAM{{\sc DiscreteJAM}}
\def\Mamposst{{\sc MAMPOSSt}}
\def\Agama{{\sc Agama}}
\def\GC{{\sc Gaia Challenge}}

\def\PyNbody{{\sc PyNbody}}

\usepackage[pdftex,
        colorlinks=true,
        urlcolor=blue,		
        filecolor=blue,		
        citecolor=blue,		
        linkcolor=blue,		
        pdftitle={},
        pdfauthor={},
        pdfsubject={},
        pdfkeywords={}
        ]{hyperref}

\def\EMCEE{{\sc emcee}}

\usepackage{array}
\newcolumntype{L}[1]{>{\raggedright\let\newline\\\arraybackslash\hspace{0pt}}m{#1}}
\newcolumntype{C}[1]{>{\centering\let\newline\\\arraybackslash\hspace{0pt}}m{#1}}
\newcolumntype{R}[1]{>{\raggedleft\let\newline\\\arraybackslash\hspace{0pt}}m{#1}}

\title[Mass modelling methods for spherical systems]{Breaking Beta: A comparison of mass modelling methods for spherical systems}

\author[J.~I.~Read et al.]{J.~I.~Read$^{1}$\thanks{E-mail: justin.inglis.read@gmail.com},  G.~A.~Mamon$^2$, E.~Vasiliev$^{3,4,5}$, L.~L.~Watkins$^{6,7,8}$, 
\newauthor
M.~G.~Walker$^9$, J.~Pe\~narrubia$^{10}$, M.~Wilkinson$^{11}$, W.~Dehnen$^{11,12}$,
P.~Das$^{4,1}$\\
$^1${\small Department of Physics, University of Surrey, Guildford, GU2 7XH, UK} \\
$^2${\small Institut d'Astrophysique de Paris (UMR 7095: CNRS \& Sorbonne Universit\'e), 75014 Paris, France} \\
$^3${\small Institute of Astronomy, University of Cambridge, Madingley Road, Cambridge, CB3 0HA, UK} \\
$^4${\small Rudolf Peierls Centre for Theoretical Physics, University of Oxford, OX1 3NP, UK} \\
$^5${\small Lebedev Physical Institute, Leninsky prospekt 53, Moscow, 119991, Russia} \\
$^6${\small AURA for the European Space Agency (ESA), ESA Office, Space Telescope Science Institute, 3700 San Martin Drive, Baltimore MD 21218, USA} \\
$^7${\small European Southern Observatory, Karl-Schwarzschild-Stra{\ss}e 2, D-85748 Garching bei M{\"u}nchen, Germany} \\
$^8${\small Department of Astrophysics, University of Vienna, T{\"u}rkenschanzstra{\ss}e 17, 1180 Vienna, Austria} \\
$^9${\small McWilliams Center for Cosmology, Department of Physics, Carnegie Mellon University,  Pittsburgh, PA 15213, USA} \\
$^{10}${\small Institute for Astronomy, University of Edinburgh, Blackford Hill, Edinburgh, EH9 3HJ, UK} \\
$^{11}${\small Department of Physics \& Astronomy, University of Leicester, Leicester, LE1 7RH, UK} \\
$^{12}${\small Universit\"ats-Sternwarte der Ludwig-Maximilians-Universit\"at, Scheinerstrasse 1, M\"unchen D-81679, Germany}
}

\begin{document}

\maketitle

\begin{abstract}
We apply four different mass modelling methods to a suite of publicly available mock data for spherical stellar systems. We focus on the recovery of the density and velocity anisotropy as a function of radius, using either line-of-sight velocity data only, or adding proper motion data. All methods perform well on isotropic and tangentially anisotropic mock data, recovering the density and velocity anisotropy within their 95\% confidence intervals over the radial range $0.25 < R/R_{1/2} < 4$, where $R_{1/2}$ is the half light radius. However, radially-anisotropic mocks are more challenging. For line-of-sight data alone, only methods that use information about the shape of the velocity distribution function are able to break the degeneracy between the density profile and the velocity anisotropy, $\beta$, to obtain an unbiased estimate of both. This shape information can be obtained through directly fitting a global phase space distribution function, by using higher order `Virial Shape Parameters', or by assuming a Gaussian velocity distribution function locally, but projecting it self-consistently along the line of sight. Including proper motion data yields further improvements, and in this case, all methods give a good recovery of both the radial density and velocity anisotropy profiles.
\end{abstract}


\section{Introduction}\label{sec:intro}

Many stellar systems are spherical or mildly triaxial, from globular clusters \citep{White&Shawl87} and tiny gas-poor dwarf spheroidal galaxies (dSphs) \citep{Sanders&Evans17}
to giant elliptical galaxies \citep{Hubble26}
and even galaxy clusters (e.g. \citealt{Limousin+13}). Building mass models of such systems has a wide range of astrophysical applications, from hunting for intermediate-mass and supermassive black holes \citep[e.g.][]{2002MNRAS.335..517V, 2008ApJ...676.1008N, 2010ApJ...710.1063V,Vitral&Mamon20} to probing the nature of dark matter \citep[e.g.][]{2011ApJ...742...20W, 2018MNRAS.481..860R, 2019MNRAS.484.1401R},
the stellar mass function \citep[e.g.][]{2012Natur.484..485C}, or the orbital velocity anisotropy \citep[e.g.][]{2019A&A...631A.131M}. The velocity anisotropy is an interesting quantity to determine as it encodes information about the assembly history of the system \citep[e.g.][]{1962ApJ...136..748E}. For spherical systems, it is typically described by the `anisotropy parameter':
\begin{equation}  \label{beta}
\beta(r) = 1 - \frac{\sigma_\theta^2(r)}{\sigma_r^2(r)} ,
\end{equation}
where $r$ is the spherically symmetric radial coordinate, $\sigma_\theta(r)$ is the tangential velocity dispersion, $\sigma_r(r)$ is the radial velocity dispersion, and $\beta(r)=1$ for radial orbits, 0 for isotropic orbits and $\beta(r) \to -\infty$ for circular orbits. 

In all of the above applications, we would like to understand how model biases and systematic errors impact the results. Mock data -- a dynamically realistic representation of the real data for which we know the true answer -- provides an elegant way to address these questions. Such mocks can be very simple, reproducing all of the assumptions employed in the analysis on real data, or highly sophisticated, simulating the effect of data selection effects and/or departures from key model assumptions. For these latter, developing the mock can be a substantial task in its own right, discouraging broad and deep tests of our methodologies.

The goal of this article is to compare how four different mass-orbit modelling algorithms, applied to mock data for a spherical, single component, dwarf galaxy, recover the radial profiles of total density and velocity anisotropy. We focus on specific physical radii, $R = [0.25,0.5,1,2,4]\,R_{1/2}$, where $R_{1/2}$ is the projected half-light radius of the tracer stars. And we consider both line-of-sight (LOS) velocity data for 1,000 and 10,000 tracer stars\footnote{We define `tracer' stars as massless particles orbiting in the combined gravitational potential of all stars, gas, dark matter (DM) etc. in a stellar system.}, and proper motion data for 1,000 stars. The line-of-sight data are similar to the state-of-the-art data that are currently available for nearby dwarf galaxies \citep[e.g.][]{2009AJ....137.3100W}, globular clusters \citep[e.g.][]{Watkins2015b,Kamann2018}, giant elliptical galaxies \citep[e.g.][]{2014MNRAS.439..659N},
and clusters of galaxies (with large spectroscopic surveys such as SDSS [\citealt{York+00}] and GAMA [\citealt{Robotham+10}]). The proper motion data are similar to what is available for globular clusters \citep[e.g.][]{Bellini2014, Watkins2015a, 2017MNRAS.468.4429Z} and what will hopefully become available for dwarf galaxies from future planned missions \citep[e.g.][]{2007ApJ...657L...1S,2017arXiv170701348T,2018NatAs...2..156M, 2020A&A...633A..36M}.

Our mock data suite is publicly available on the {{\sc Gaia Challenge}} wiki site\footnote{\label{surrey}\href{http://astrowiki.ph.surrey.ac.uk/dokuwiki/}{http://astrowiki.ph.surrey.ac.uk/dokuwiki/} (under the `Spherical \& Triaxial' suite tab).} and has already been used to test a number of methods in the literature \citep{2011ApJ...742...20W,2015PhRvD..91h3535G, 2015MNRAS.453..849B, 2014MNRAS.441.1584R, 2016MNRAS.463.1117Z, 2016MNRAS.463.3630G, 2017arXiv170104833R, 2017MNRAS.466..669C, 2019MNRAS.482.3356D}. These mock data will continue to grow and improve in sophistication over time. Our central philosophy is that {\it it is much easier to produce a sophisticated mock, than to produce a sophisticated mass modelling tool.} Using such mocks may demonstrate that simple models -- e.g. those that assume spherical symmetry -- work sufficiently well to test or rule out interesting models, or estimate a given parameter of interest. In this case, increasing the sophistication of the model is not required. Mock data tests can also help us identify problems in the real data -- for example, if confidence intervals on the model are much smaller than expected from the mock. 

In this paper, we assess the performance of four mass modelling methods described in \S\ref{sec:methods} through their recovery of the radial density and velocity anisotropy profiles. Three of these methods, \GravSphere, \Mamposst, and \DiscreteJAM, are based on the Jeans (\citeyear{1922MNRAS..82..122J}) equation of local dynamical equilibrium, while one, \Agama, is based on fitting a global distribution function. For stationary spherical systems without streaming motions, the Jeans equation is given by \citep[e.g.][]{2008gady.book.....B}:
\begin{equation}  \label{jeans}
\frac{\dd \left(\nu(r) \sigma_r^2(r)\right)}{\dd r} + 2\,\frac{\beta(r)}{r}\,\nu(r)\,\sigma_r^2(r) = -\nu(r) \,\frac{G\,M(r)}{r^2} \ ,
\end{equation}
where $\nu(r)$ is the tracer density, $M(r)$ is the total enclosed mass, $\beta(r)$ is the anisotropy parameter (\ref{beta}), and $G$ is Newton's gravitational constant. Jeans methods have the advantage that they can be efficiently implemented, allowing a wide range of models to be explored \citep[e.g.][]{2017arXiv170104833R}. Furthermore, since they do not require any assumption about the form of the distribution function, they can be constructed to be formally unbiased \citep[e.g.][]{2014JPhG...41f3101R,2017arXiv170104833R}. However, disadvantages include the fact that the fits can, in principle, require a formally negative (i.e. unphysical) distribution function \citep[e.g.][]{2006ApJ...642..752A}. By contrast, methods like \Agama\ have the advantage that they provide a full distribution function that is positive definite and that can be easily convolved with errors, binary contamination \citep[e.g.][]{2010ApJ...722L.209M, 2018AJ....156..257S} and survey selection functions. However, if the chosen parameterisation of the distribution function does not contain the true solution, such methods run the risk of becoming biased \citep[e.g.][]{2011MNRAS.416.2318G}. 

We focus here on a suite of spherical, idealised, mocks with minimal measurement uncertainties as a starting point. As we shall see, already this presents a challenge for some methods, primarily due to the mass-anisotropy degeneracy \citep[e.g.][]{Binney&Mamon82, 2017arXiv170104833R}. More sophisticated mocks that break spherical symmetry and include the impact of unbound tidally stripped stars, lower numbers of tracer stars, and multiple independent tracer populations are also available through the \GC\ wiki site. These have already been presented in the literature for the \GravSphere\ method \citep{2017arXiv170104833R, 2018MNRAS.481..860R, 2019MNRAS.485.2010G}, with the result that -- at least for 1,000 tracers with small velocity uncertainties -- these additional complications do not dominate the error budget. Other groups have reached similar conclusions with their own mock data tests, including on cosmologically realistic mocks \citep[e.g.][]{Sanchis+04,2009MNRAS.399..812W,Mamon+13,2013MNRAS.431.2796K, 2020arXiv200713780H, 2019arXiv191109124G}. We will consider such more sophisticated tests on the methods presented here in future work.

This paper is organised as follows. In \S\ref{sec:mocks}, we describe the mock data suite and how it was set up. In \S\ref{sec:methods}, we describe the different mass modelling techniques we apply to these mock data. In \S\ref{sec:results}, we apply several modelling methods to the spherical mocks to examine the accuracy of their inferred density $\rho(r)$ and velocity anisotropy $\beta(r)$ profiles. Finally, in \S\ref{sec:conclusions}, we discuss our results and present our conclusions.

\section{Mock data}\label{sec:mocks}

\begin{table*}
\begin{center}
\begin{tabular}{l | cccc}
Label & $[r_*,\alpha_*,\beta_*,\gamma_*]$ & $[(\rho_0,r_\DM,\alpha_\DM,\beta_\DM,\gamma_\DM)]$ & $\beta_0,\beta_\infty,n,r_a$ & $R_{1/2}$ \\
\hline
\rowcolor{gray!40} PlumCuspIso & $[0.25,2,5,0.1]$ & $[6.4,1,1,3,1]$ & $[0,0,1,0]$ & 0.25 \\
PlumCoreIso & $[1,2,5,0.1]$ & $[40,1,1,3,0]$ & $[0,0,1,0]$ & 1.0 \\
\rowcolor{gray!40} NonPlumCuspIso & $[0.25,2,5,1]$ & $[6.4,1,1,3,1]$ & $[0,0,1,0]$ & 0.20  \\
NonPlumCoreIso & $[1,2,5,0.1]$ & $[40,1,1,3,0]$ & $[0,0,1,0]$ & 0.79  \\
\rowcolor{gray!40} PlumCuspOM & $[0.1,2,5,0.1]$ & $[6.4,1,1,3,1]$ & $[0,1,2,0.10]$ & 0.1  \\
PlumCoreOM & $[0.25,2,5,0.1]$ & $[40,1,1,3,0]$ & $[0,1,2,0.25]$ & 0.25  \\
\rowcolor{gray!40} NonPlumCuspOM & $[0.1,2,5,1]$ & $[6.4,1,1,3,1]$ & $[0,1,2,0.10]$ & 0.079  \\
NonPlumCoreOM & $[0.25,2,5,1]$ & $[40,1,1,3,0]$ & $[0,1,2,0.25]$ & 0.20  \\
\rowcolor{gray!40} PlumCuspTan & $[0.5,0.5,5,0.1]$ & $[2.39,2,1,4,1]$ & $[-0.5,-0.5,2,0]$ & 0.86  \\
PlumCoreTan & $[1.75,0.5,5,0.1]$ & $[3.0,4.0,1,4,0]$ & $[-0.5,-0.5,2,0]$ & 3.0  \\
\rowcolor{gray!40} NonPlumCuspTan & $[0.5,0.5,5,1.0]$ & $[2.39,2,1,4,1]$ & $[-0.5,-0.5,2,0]$ & 0.41  \\
NonPlumCoreTan & $[1.75,0.5,5,1.0]$ & $[3.0,4.0,1,4,0]$ & $[-0.5,-0.5,2,0]$ & 1.43  \\
\hline
\end{tabular}
\end{center}
\caption{
Parameters of the spherical mock data that we use in this paper. The columns give, from left to right: the mock data label; the tracer density parameters (see equation \ref{eqn:zhao}); the DM halo density parameters (see equation \ref{eqn:zhao}); the velocity anisotropy profile; and the projected half stellar mass radius. The length units are given in kpc, while $\rho_0$ has units of $10^7$\,M$_\odot$\,kpc$^{-3}$. The mock labelling convention is: $<$light profile$>$$<$Cusp/core$>$$<$Anisotropy$>$, where: the $<$light profile$>$ can be Plummer-like (Plum; $\gamma_* = 0.1$) or cusped (NonPlum; $\gamma_* = 1$); the DM halo can be cusped (Cusp; $\gamma_\DM = 1$) or cored (Core; $\gamma_\DM = 0$); and the $<$Anisotropy$>$ can be isotropic (Iso; $\beta = 0$), of generalized Osipkov-Merritt form (OM; \citealt{1979PAZh....5...77O, Merritt85}, see equation \ref{eqn:betaprofile}) or tangentially anisotropic (Tan; $\beta = -0.5$).
}
\label{tab:mocks}
\end{table*}

In this section, we describe our mock data; the full `default spherical' suite is summarised in Table \ref{tab:mocks} and is publicly available at the \GC\ wiki site$^{\ref{surrey}}$. All of the mock data have a single population of massless tracer stars orbiting within a host DM halo. They are available either without errors or with normally distributed errors with a standard deviation of $2$\,km\,s$^{-1}$ on the line-of-sight velocities for each star (larger errors can easily be added to explore their effect, but in this paper we will focus on the case where the measurement uncertainties do not dominate over the sampling error). We also include proper motion data, assuming similarly small measurement errors.

As discussed in \S\ref{sec:intro}, the Gaia Challenge suite also includes a host of more realistic triaxial, tidally stripped and split-population mocks. We will consider more realistic mocks with individual star-by-star velocity errors, binary star contamination, contamination from Milky Way stars, unbound tidally stripped stars, and a realistic survey selection function, in future work. We note, however, that -- at least for mock galaxies similar to the Milky Way `classical' dwarfs -- previous works have found these additional effects to be small \citep[e.g.][]{2011ApJ...742...20W, 2018MNRAS.481..860R, 2019arXiv191109124G}.

We label the mocks using the following naming convention:

\vspace{-2mm}
\begin{equation}
<{\rm light \; profile}><{\rm DM\; cusp/core}><{\rm Anisotropy}> \nonumber
\end{equation}
where: the $<$light profile$>$ can be Plummer-like (\citealt{1911MNRAS..71..460P}; Plum) or cusped (NonPlum); the DM halo can be cusped (Cusp) or cored (Core); and the $<$Anisotropy$>$ can be isotropic (Iso), of generalized Osipkov-Merritt form (OM; see next section) or tangentially anisotropic (Tan). 

The isotropic and radially anisotropic spherical mocks are a subset of the mocks presented in \citet{2011ApJ...742...20W} (see \S\ref{sec:radialmocks}); the tangentially anisotropic mocks were generated using the method outlined in \citet{2009MNRAS.395.1079D} (see \S\ref{sec:tangentialmocks}). In all cases, we assume a double-power-law profile \citep{1996MNRAS.278..488Z} for both the tracer distribution and the host DM halo:
\begin{equation}
  \rho_X(r) = \rho_0\left(\frac{r}{r_X}\right)^{-\gamma_X} \left[1+\left(\frac{r}{r_X}\right)^{\alpha_X}\right]^{(\gamma_X-\beta_X)/\alpha_X}
  \label{eqn:zhao}
\end{equation}
where $\rho_0$ is a normalisation parameter; $r_X$ sets the scale length; $\gamma_X$ is the inner asymptotic logarithmic slope; $\beta_X$ is the outer asymptotic logarithmic slope; and $\alpha_X$ controls the sharpness of the transition at $r_X$. For the tracers, we use the notation: $\nu = \rho_*(r;\;\nu_0,r_*,\alpha_*,\beta_*,\gamma_*)$; for the DM we write similarly: $\rho_\DM = \rho_\DM(r;\;\rho_0,r_\DM,\alpha_\DM,\beta_\DM,\gamma_\DM)$.

\subsection{Radially anisotropic mocks}\label{sec:radialmocks}

The velocity anisotropy coefficient (\ref{beta}) is assumed to depend on radius as follows:
\begin{equation} 
\beta(r) = \frac{ \beta_\infty r^n + \beta_0 r_a^n }{ r^n + r_a^n },
\label{eqn:betaprofile}
\end{equation}
where $\beta_0$ is the asymptotic anisotropy at small radii, $\beta_\infty$ is the asymptotic anisotropy at large radii, $r_a$ is the transition radius, and $n$ controls the sharpness of the transition. A familiar special case is the Osipkov--Merritt anisotropy profile \citep{1979PAZh....5...77O, Merritt85}, used already above for the mock data, in which $\beta_0 = 0$, $\beta_\infty = 1$, and $n = 2$.

For the radially-anisotropic mocks, the tracers are set up in equilibrium inside their host DM halo assuming an OM distribution function, as in \citet{2011ApJ...742...20W}.

\subsection{Tangentially anisotropic  mocks}\label{sec:tangentialmocks}

The tangentially-anisotropic mocks are set up using the `made to measure' code from \citet{2009MNRAS.395.1079D} that is an evolution of the method described in \citet{1996MNRAS.282..223S}. In such methods, an $N$-body system is evolved for some time under its own self-gravity, adjusting the masses of each particle so as to move towards a target phase space distribution. The tangentially anisotropic mocks also assume the form given in equation \ref{eqn:zhao} for the tracer and DM density profiles. 

Tangential anisotropy is not expected theoretically,
unless significant angular momentum is imparted to the system through mergers, or the system is orbiting within a strong tidal field\footnote{Note that in the case of strong tides, such tangential anisotropy would indicate a departure from the steady-state pseudo-equilibrium assumed by all of the methods explored here.}  \citep[e.g.][]{2006MNRAS.366..429R}. Given the lack of any strong theoretical motivation for a particular form of tangential anisotropy, we assume a constant $\beta = -0.5$ at all radii to explore whether tangential mocks can or cannot be successfully recovered.

\section{Methods}\label{sec:methods}

In this section, we briefly describe each of the four mass-orbit modelling methods that will be applied to the mock data.

\subsection{\GravSphere}
\label{section:GravSphere}

\GravSphere\footnote{A public version of the code, {{\sc PyGravSphere}}, can be downloaded from \href{https://github.com/AnnaGenina/pyGravSphere}{https://github.com/AnnaGenina/pyGravSphere}.} is described and tested in detail in \citet{2017arXiv170104833R}, \citet{2018MNRAS.481..860R} and \citet{2019arXiv191109124G}. It fits the line-of-sight velocity variance, which can be written by combining the Jeans equation~(\ref{jeans}) with an equation projecting the velocities along the line of sight, yielding \citep{Binney&Mamon82}:
\begin{equation}
\frac{\Sigma(R)\,\siglos^2(R)}{2}= \int_R^\infty \left(1\!-\!\beta\frac{R^2}{r^2}\right)
    \nu\sigma_r^2\,\frac{r\,\dd r}{\sqrt{r^2\!-\!R^2}} \ ,
    \label{eqn:LOS}
\end{equation}
where $\Sigma(R)$ denotes the tracer surface mass density at projected radius $R$. 

The radial velocity variance is given by (\citealt{vanderMarel94}, in this form by \citealt{Mamon&Lokas05}):

\begin{equation}
\sigma_r^2(r) = \frac{1}{\nu(r) g(r)} \int_r^\infty \frac{GM(\tilde{r})\nu(\tilde{r})}{\tilde{r}^2} g(\tilde{r}) \dd \tilde{r}
\label{eqn:main}
\end{equation}
where
\begin{equation}
g(r) = \exp\left(2\int_0^r \frac{\beta(\tilde r)}{\tilde r}\dd \tilde r\right)
\label{eqn:ffunc}
\end{equation}
and $M(r)$ is the cumulative mass of the stellar system (due to all stars, gas, DM etc.).

\GravSphere\ uses a free-form, i.e. non-parametric, model for $M(r)$ that comprises a contribution from all visible matter and a contribution from `dark matter' that is described by a sequence of power laws defined on a set of radial bins. By default, these bins are defined at $R=[0.25,0.5,1,2,4]\,R_{1/2}$ where $R_{1/2}$ is the projected half light radius of the tracer stars. (This can be changed if the data quality warrant it, but in this paper we will use only this default choice.) The tracer light profile is also non-parametric, using a series sum of Plummer spheres, as in \citet{2016MNRAS.459.3349R}. This has the advantage that the mapping between the spherically-averaged tracer density, $\nu(r)$ and the projected light profile, $\Sigma_*$ is analytic \citep{2017arXiv170104833R}. Finally, the velocity anisotropy, $\beta(r)$, is assumed to have a form (\ref{eqn:betaprofile}) that ensures broad generality while making the function $g(r)$ (equation \ref{eqn:ffunc}) analytic. For assigning priors on the parameters in equation \ref{eqn:betaprofile}, we transform $\beta(r)$ to a symmetrised velocity anisotropy, as in \citet{2006MNRAS.tmp..153R} and \citet{2017arXiv170104833R}:
\begin{equation} 
\betastar =  \frac{\sigma_r^2 - \sigma_t^2}{\sigma_r^2 + \sigma_t^2} = \frac{\beta}{2-\beta}
\label{eqn:betastar}
\end{equation}
This is bounded on $-1 < \betastar < 1$ ($\betastar = -1$ corresponds to full tangential anisotropy; $\betastar = 1$ to full radial anisotropy; and $\betastar = 0$ to isotropy), unlike $\beta$ that tends to $-\infty$ for full tangential anisotropy.

\GravSphere\ can also use split population or proper motion data, where available. By default, it also fits the two higher order `Virial Shape Parameters' (VSPs; \citealt{1990AJ.....99.1548M, 2014MNRAS.441.1584R}):
\begin{eqnarray} 
\vsone & = & \frac{2}{5}\, \int_0^{\infty} G M\, (5-2\beta) \,\nu \sigma_r^2 \,r \,\dd r \\
\label{eqn:vs1}
& = & \int_0^{\infty} \Sigma \vlosfour\, R\, \dd R
\label{eqn:vs1data}
\end{eqnarray}
and:
\begin{eqnarray} 
\vstwo & = & \frac{4}{35} \,\int_0^{\infty} G M\, (7-6\beta)\, \nu \sigma_r^2 \,r^3 \,\dd r \\
\label{eqn:vs2}
& = & \int_0^{\infty} \Sigma \vlosfour\, R^3\, \dd R \ .
\label{eqn:vs2data}
\end{eqnarray}
The key advantage of these VSPs is that they involve fourth-order moments of the line-of-sight velocities $\vlosfour$, but depend only on $\beta(r)$ and not on its fourth-order counterparts \citep{1990AJ.....99.1548M,2014MNRAS.441.1584R,2017arXiv170104833R}. Thus, $\vsone$ and $\vstwo$ allow us to obtain additional constraints on $\beta(r)$ via the line-of-sight velocities, alleviating the mass-anisotropy degeneracy between the spherically averaged density profile $\rho(r)$ and the velocity anisotropy $\beta(r)$.

The errors on $\siglos$, $\vlosfour$, $\vsone$ and $\vstwo$ are estimated as in \citet{2018MNRAS.481..860R}, using a Monte-Carlo method that incorporates the individual measurement errors on each star.

Finally, if proper motion data are available, \GravSphere\ can fit also for the radial and tangential plane-of-sky velocity dispersions, $\sigma_{\rm POSr}$ and $\sigma_{\rm POSt}$, satisfying \citep[e.g.][]{2007ApJ...657L...1S,2010ApJ...710.1063V,2017arXiv170104833R}:

\begin{eqnarray}
\frac{\Sigma(R)\,\sigma_{\rm POSr}^2(R)}{2} &\!\!\!\!=\!\!\!\!& 
\int_R^\infty \left(1-\beta+\beta\frac{R^2}{r^2}\right)
    \frac{\nu(r)\,\sigma_r^2 r\,\dd r}{\sqrt{r^2-R^2}} \ ,
    \label{eqn:sigpmr} \\
\frac{\Sigma(R)\,\sigma_{\rm POSt}^2(R)}{2}
&\!\!\!\!=\!\!\!\!& 
\int_R^\infty \left(1-\beta\right)
    \frac{\nu(r)\,\sigma_r^2 r\,\dd r}{\sqrt{r^2-R^2}}\ .
    \label{eqn:sigpmt}
\end{eqnarray}
This provides an alternative route to breaking the mass-anisotropy degeneracy since equations~(\ref{eqn:LOS}), (\ref{eqn:sigpmr}) and (\ref{eqn:sigpmt}) each have different dependencies on $\beta$ (see \citealt{2017arXiv170104833R} for further details).

\GravSphere\ fits the above model to the surface density profile of tracer stars, $\Sigma_*(R)$, their line-of-sight projected velocity dispersion profile $\siglos(R)$ and their VSPs using the \EMCEE\ affine invariant Markov Chain Monte Carlo (MCMC) sampler from \citet{2013PASP..125..306F}. We assume uncorrelated Gaussian errors such that the Likelihood function is given by $\mathcal{L} = \exp(-\chi^2/2)$. We use as default 1,000 walkers, each generating 5,000 models and we throw out the first half of these as a conservative `burn in' criteria. Please see \citet{2017arXiv170104833R} for further details of our likelihood function, methodology and priors. 

\subsection{\DiscreteJAM}

The \DiscreteJAM\ method uses an alternative approach to solve the Jeans equations, which were laid out in \autoref{section:GravSphere}. There are three key features of this method: (1) both the tracer number density profile $\nu(r)$ and the underlying mass density profile $\rho(r)$  are parametrised in the form of Multi-Gaussian Expansions \citep[MGEs,][]{Emsellem1994}; (2) we fit to the discrete position measurements for each star, not to a binned tracer-density profile; and (3) we fit to the discrete velocity measurements for each star, not to a binned velocity-dispersion profile.

We model the dwarf galaxies using the Jeans Anisotropic MGE (JAM) models \citep{Cappellari2008, Watkins2013, Cappellari2015}. For the spherical mocks that are the focus of this paper, we use the spherical JAM models, which assume spherical symmetry and allow for anisotropy (defined in spherical coordinates) but do not include rotation.

An MGE is characterised by a set of Gaussian components, each with a width and a weight. If the tracer distribution is made up of $N_\mathrm{t}$ components where each component $j$ has a weight $\nu_j$ and a width $s_j$, then the tracer density profile of the MGE is:
\begin{equation}
    \nu \left( r \right) = \sum_{j=1}^{N_\mathrm{t}} \nu_j \exp \left( -\frac{r^2}{2 s_j} \right).
	\label{equation:tracer_mge}
\end{equation}
Similarly, the mass density $\rho(r)$ can be parametrised by a set of $N_\mathrm{m}$ Gaussians with each component $k$ having weight $\rho_k$ and width $s_k$. The density profiles can be fit analytically (where we fit MGEs to assumed functional forms and fit for the parameters of the functional forms) or non-parametrically (where we fit the MGE properties directly). Here we choose an analytic approach, but see \citet{HenaultBrunet2019} for a non-parametric methodology.

We assume that both the tracer density and DM density are described by \autoref{eqn:zhao}. The models are insensitive to the normalisation of the tracer density $\nu_0$, so this leaves us with four free parameters relating to the tracer distribution $(r_\star, \alpha_\star, \beta_\star, \gamma_\star)$, and five relating to the DM distribution $(\rho_0, r_\DM, \alpha_\DM, \beta_\DM, \gamma_\DM)$. We assume that the anisotropy profile is of the generalized OM form (\autoref{eqn:betaprofile}) with fixed $n = 2$. As in the \GravSphere\ method, we symmetrise $\beta_0$ and $\beta_\infty$ using \autoref{eqn:betastar} to avoid the high anti-symmetry and infinite lower bound for $\beta$. This leaves three free components for the anisotropy: $(r_a, \betastar_0, \betastar_\infty)$.

We sample scale density $\rho_0$ and scale radii $r_\star$, $r_\DM$ and $r_a$ in log-space as negative values would be unphysical, these parameters have the potential to explore many orders of magnitude in size when model fitting, and a dimensionless parameter space is generally more efficient to search. The other parameters are already dimensionless by definition.

We use an unbounded flat prior on $\log \rho_0$. We restrict the range of the tracer scale radii to $-2 \le \log_{10} (r_\star/\mathrm{kpc}) \le 1$ and the DM scale radii to $0.25 \le r_\DM/r_\star \le 50$, but otherwise assume flat priors within these ranges. For transition parameters $\alpha_\star$ and $\alpha_\DM$, we use a flat prior in $0.5 \le (\alpha_\star, \alpha_\DM) \le 2.5$, this eliminates models that change too sharply or too slowly.

For the outer density slopes, we use flat priors in $2 \le \beta_\star \le 7$ and $2 \le \beta_\DM \le 5$, where the upper limits restrict the steepness based on observed values. Models with$(\beta_\star,\beta_\DM) \le 3$ are unphysical as they are not finite, however, some mocks have outer slopes of 3, meaning that the correct answer is at the edge of the parameter space. To allow the fits to fully explore the parameter space, we use lower limits of 2 for the outer slopes.

For the inner density slopes, we use flat priors in $-0.5 \le \gamma_\star \le \beta_\star$ and $-0.5 \le \gamma_\DM \le \beta_\DM$, where the upper limits ensure that the inner slope is shallower than the outer slope. Formally, a lower limit of 0 on the inner slopes we would be appropriate here, otherwise we have a negative density component in the MGE fit. However, as above, some mocks have inner slopes of 0, thus at the edge of the parameter space. So to allow the models to fully explore the region of parameter space around the true value, we use lower limits of $-0.5$ on the inner slopes.

We assume a flat prior on the anisotropy scale radii in $0.5 \le r_a/r_\star \le 2$; both the isotropic and tangential mocks have flat anisotropy profiles in which case the notion of a anisotropy transition radius becomes somewhat meaningless, with this choice of prior we give the models some flexibility, but avoid fitting a potentially undefined parameter. In theory, $-1 \le \betastar \le 1$. In practice, we set a lower bound of $\beta = -40$, which corresponds to $\betastar \sim -0.95$ as extremely tangential models with $\betastar \sim -1$ are computationally challenging (and physically unlikely). We use a flat prior on $\betastar$ within these limits.

For a given set of density parameters, we calculate $\nu(r)/\nu_0$ or $\rho(r)$ using \autoref{eqn:zhao}. To this we fit an MGE using \textsc{mgefit} \citep{Cappellari2002}, which we then deproject to obtain the projected surface mass and tracer density profiles as needed for the JAM models.\footnote{A python package \textsc{mgetools} for manipulating MGEs is available at \url{http://www.github.com/lauralwatkins/mgetools}.} To each component $j$ of the tracer MGE we assign an anisotropy $\beta_j$ by calculating the value of the generalized OM profile at the radius of each Gaussian component $s_j$.

Our observables are the spatial distribution of the tracer particles and the velocity measurements of the tracer particles. We fit to both discretely, that is we do not bin the tracers but consider each star individually. We follow the approach laid out in \citet{Watkins2013}, modified as described below. This method is flexible and can be used to fit only LOS velocity information, only proper motions, or full 3D velocity information. The original method was designed to fit a binned surface brightness profile, whereas here we fit the tracer density discretely as well. 

Consider a tracer particle $i$ at a projected distance $R_i$ from the centre of the dwarf with velocity $v_{X,i} \pm \delta_{X,i}$, where X can be the Right Ascension PM ($\alpha$), the Declination PM ($\delta$) or the LOS ($z$), and the uncertainties $\delta_{X,i}$ are assumed to be Gaussian. To fit the spatial distribution of the tracers, we use the projected tracer density MGE to calculate the normalised probability $P_i = P(R_i)$ of observing a star at projected radius $R_i$. Then the likelihood of all $N$ tracer particles is:
\begin{equation}
    \mathcal{L}_\star = \prod_{i=1}^N P_i .
\end{equation}
To fit the velocity measurements, we assume that the LOS velocity distribution at a given radius is Gaussian with mean velocity 0 (due to the assumptions of spherical symmetry and no rotation) and velocity dispersion $\sigma_{X,i}$, which we calculate from the JAM model. Then the likelihood of all $N$ measurements is:
\begin{equation}
	\mathcal{L}_\mathrm{kin,X} = \prod_{i=1}^N \frac{1}{\sqrt{2 \pi \left( \delta_{X,i}^2 + \sigma_{X,i}^2 \right)}} \exp \left( \frac{-v_{X,i}^2}{2 \left( \delta_{X,i}^2 + \sigma_{X,i}^2 \right)} \right) ,
\end{equation}
which accounts for the fact that the observed dispersion is a convolution of the true dispersion $\sigma_{X,i}$ and the measurement uncertainties $\delta_{X,i}$. When using only LOS velocity information, the total likelihood $\mathcal{L}$ for the spatial and kinematic distributions combined is then:
\begin{equation}
	\mathcal{L} = \mathcal{L}_\star \mathcal{L}_\mathrm{kin,z}.
\end{equation}
When using PM and LOS kinematic information, the total likelihood $\mathcal{L}$ becomes:
\begin{equation}
	\mathcal{L} = \mathcal{L}_\star \mathcal{L}_\mathrm{kin,\alpha}
	    \mathcal{L}_\mathrm{kin,\delta} \mathcal{L}_\mathrm{kin,z},
\end{equation}
assuming that the measurements and uncertainties are not correlated.

The posterior probability of the model given the data is then obtained by multiplying together the likelihood and parameter priors. By maximising the posterior, we can locate the family of models that best fit the data. To efficiently explore our parameter space and locate the high-posterior region, we use \EMCEE\ \citep{2013PASP..125..306F} with 100 walkers and for 20,000 steps.

\subsection{MAMPOSSt}

\Mamposst\ \citep*{Mamon+13} is similar to \DiscreteJAM, as it fits the full distribution of the  individual tracers in projected phase space. But while \DiscreteJAM\ assumes that the line-of-sight velocities are Gaussian-distributed, \Mamposst\ avoids this assumption, since anisotropy affects the shape of the line-of-sight velocity distribution \citep{Merritt87}, and assumes instead that the local three-dimensional velocity distribution is Gaussian-distributed. More precisely, \Mamposst\ assumes parametric forms for the mass, tracer density and anisotropy profiles. In situations where only line-of-sight velocities are available,  the density of tracers in projected phase space (projected radius $R$ and the line-of-sight velocity $v_{\rm LOS}$) is the integral of the local velocity distributions along the line of sight: 
\begin{equation}
\!\!\!\!g(R, v_{\rm LOS}) = 
2\,
\int_R^\infty
h\left (v_{\rm LOS}|R,r\right)\,\nu(r)\,\frac{r\,\dd r}{\sqrt{r^2\!-\!R^2}}
\,,
\end{equation}
\Mamposst\ assumes that
 the {\it local} velocity distribution function $h$ is a  Gaussian of zero mean:
 \begin{equation}
  h(v_{\rm LOS}\,|\,R,r) =   {\cal G}\left(v_{\rm LOS}, 0, \sigma_{\rm LOS}^2\right) \ ,
 \end{equation}
 where
$\sigma_{\rm LOS}^2(R,r) = [1-\beta(r) R^2/r^2]\,\sigma_r^2$
 as in the integrand of equation~(\ref{eqn:LOS}).
The likelihood is
given by:
\begin{equation}
-\ln {\cal L}=-\sum_i \ln q(R_i,v_i\,|\,\bm{\theta}) \ ,
\end{equation}
where $i$ are the indices of the individual tracers, $v_i$ is the LOS velocity for tracer particle $i$, $\bm{\theta}$ is the vector of parameters, while $q$ is the probability density:
\begin{equation}
q(R,v)=\frac{2\pi\,R\,g(R,v)}{N_{\rm p}\left(R_{\rm max}\right)-N_{\rm p}\left(R_{\rm min}\right)} \ ,
\end{equation}
where $N_{\rm p}(R) = \int_0^R 2\pi R'\, \Sigma(R')\,{\rm d}R'$ is the predicted number of tracers expected within the projected radius $R$. 

\Mamposst\ has recently been generalized to include handling of proper motion data (\Mamposst-PM, Mamon \& Vitral, in prep.). The full local velocity distribution function is assumed to be the product of three Gaussians:
\begin{eqnarray}
    h\left ({\bf v}|R,r\right)  &\!\!\!\!=\!\!\!\!& 
    {\cal G}\left(v_{\rm LOS}, 0, \sigma_{\rm LOS}^2(R,r)\right) \nonumber \\
    &\!\!\!\!\mbox{}\!\!\!\!& \times
    {\cal G}\left(v_{\rm POSr}, 0, \sigma_{\rm POSr}^2(R,r)\right) \,\nonumber \\
        &\!\!\!\!\mbox{}\!\!\!\!& \times
    {\cal G}\left(v_{\rm POSt}, 0, \sigma_{\rm POSt}^2(R,r)\right) \ ,
\end{eqnarray}
where the 
plane-of-sky velocity variances are $\sigma_{\rm POSr}^2(R,r)=[1-\beta(r)+\beta(r)R^2/r^2]\,\sigma_r^2(r)$ and $\sigma_{\rm POSt}^2(R,r)=[1-\beta(r)]\,\sigma_r^2(r)$ as given in the integrands of equations~(\ref{eqn:sigpmr}) and (\ref{eqn:sigpmt}), respectively, and where
$\sigma_r^2$ is the solution of the Jeans equation taken from equations~(\ref{eqn:main}) and (\ref{eqn:ffunc}).

The likelihood is now
given by:
\begin{equation}
-\ln {\cal L}=-\sum_i \ln q(R_i,{\bf v}_i\,|\,\bm{\theta}) \ ,
\end{equation}
where $i$ are the indices of the individual tracers, ${\bf v}_i$ is the  velocity vector for tracer particle $i$, $\bm{\theta}$ is the vector of parameters, while the probability density $q$ is now
\begin{equation}
q(R,{\bf v})=\frac{2\pi\,R\,g(R,{\bf v})}{N_{\rm p}\left(R_{\rm max}\right)-N_{\rm p}\left(R_{\rm min}\right)} \ .
\end{equation}

For both \Mamposst\ and \Mamposst-PM, measurement uncertainties are added in quadrature to the predicted line-of-sight and plane-of-sky velocity dispersions in the local velocity distribution function $h$.

The mass profile is assumed to follow Equation~(\ref{eqn:zhao}), for which the 
enclosed total mass (here assuming only dark matter contributes) is
\begin{eqnarray}
 M(r) &\!\!\!\!=\!\!\!\!&
 \frac{4\pi}{3-\gamma}\,
\rho_0 \,r^3\left(\frac{r_{D}}{r}\right)^\gamma 
\nonumber \\
&\!\!\!\!\mbox{}\!\!\!\!&\times \ {}_2F_1\left[
\frac{3\!-\!\gamma}{\alpha},
\frac{\beta\!-\!\gamma}{\alpha},
\frac{3\!-\!\gamma}{\alpha}+1,
-\left(\frac{r}{r_{D}}\right)^{\alpha}\right] \,,
\end{eqnarray}
where $\alpha=\alpha_{D}$,
$\beta=\beta_{D}$,
$\gamma=\gamma_{D}$, and
${}_2F_1$ is the ordinary hypergeometric function.
The \Mamposst\ analyses assume 
$\alpha_{\rm D}=1$ and 
 free inner and outer slopes.
 The tracer density profile is also described by Equation~(\ref{eqn:zhao}), with $\beta_*=5, \alpha_*=2$ and a free inner slope $\gamma_*$. Finally, the velocity anisotropy profile has the form of Equation (\ref{eqn:betaprofile}) with $n=2$ and free $\beta_0$, $\beta_\infty$ and $r_a$. 

The nine free parameters are thus the mass normalisation, the mass scale radius $r_2$, the DM inner and outer slopes $\gamma_{\rm D}, \beta_{\rm D}$, the stellar scale radius $r_*$ and inner density slope $\gamma_*$, the inner and outer anisotropies $\beta_0$ and $\beta_\infty$, and the anisotropy radius $r_a$.

The minimisation is performed with the CosmoMC \citep{Lewis&Bridle02} MCMC sampler, using 6 chains with 90\,000 elements, discarding the first 15\,000 elements of each chain for the analysis.
 We adopted $R_{\rm min}=0$ and $R_{\rm max}=1.9\,\rm kpc$ for our allowed range of projected radii.
Flat priors were used on the log masses, log scales, and on the indices and symmetrised anisotropies, with ranges listed in Table~\ref{tab:assump}.

\subsection{DF fitting with the \Agama\ library}

The \Agama\ library \citep{Vasiliev19} is a C++ library for constructing galaxy models using distribution functions (DF). A DF, $f(\boldsymbol{w})$, fully describes the distribution of a population of stars with phase-space coordinates, $\boldsymbol{w}\equiv\{\boldsymbol{x},\boldsymbol{v}\}$. In a steady state, $f$ only depends on the integrals of motion, $\boldsymbol{I}(\boldsymbol{w})$, supported by the potential. The log-likelihood of a model $\mathcal{M}$ characterised by the potential $\Phi(r)$ and the DF $f(\boldsymbol{I})$ is given by the sum of log-probabilities of each star being drawn from this DF (if we assume that all stars could be observed, i.e., the selection function is unity):
\begin{equation}
\ln\mathcal{L} = \sum_{i=1}^{N_\mathrm{stars}} \ln f\big[\boldsymbol{I}(\boldsymbol{w}_i;\, \Phi)\big].
\end{equation}

In the case of incomplete data we need to marginalise over the probability of each star
by integrating over missing phase-space coordinates. In the case of errors, we need to convolve with the distribution of errors.
For instance, if we have only the projected radius $R$ and the line-of-sight velocity $v_z$
that carries a Gaussian error $\delta v_z$, the above expression is modified to
\begin{align}  \label{eq:MarginalizedLikelihoodDF}
\ln\mathcal{L} &= \sum_{i=1}^{N_\mathrm{stars}} \ln
\Bigg\{ \int_{-\infty}^\infty dz \int_{-\infty}^\infty dv_x \int_{-\infty}^\infty dv_y
\int_0^1 d\xi\,\times \nonumber\\
 &  f\Big[\boldsymbol{I}(R,z,v_x,v_y,v_z+\sqrt{2}\,\delta v_z\,\mathrm{erf}^{-1}(2\xi-1);\,\Phi)\Big]
\Bigg\},
\end{align}
where $\xi\sim \mathcal{U}(0,1)$ marginalises over velocity error.

We evaluate these multidimensional integrals by using the Monte Carlo approach. For each star with index $i$, we construct an array of $N_\mathrm{samp}\gtrsim 10^3$ ``sample points'' with missing phase-space coordinates $\boldsymbol{w}_{i,k}$ assigned from a suitable prior distribution $P_i(\boldsymbol{w})$. The marginalised value of DF for the $i$-th star is $f_i = \sum_{k=1}^{N_\mathrm{samp}} f\big[\boldsymbol{I}(\boldsymbol{w}_{i,k};\, \Phi)\big] / P_i(\boldsymbol{w}_{i,k})$. Some of these samples would have a zero probability (e.g., if the energy is positive), but as long as there remain samples in a valid region of phase space, the total probability is positive, and hence the log-likelihood is finite. 

To eliminate the impact of Poisson noise on the relative odds of different models, we use the same set of sample points for all models \citep[e.g.][]{2013MNRAS.433.1411M}, and to further improve the accuracy, we design the priors according to the importance sampling approach, i.e. we place samples more densely in regions that are expected to have a higher value of $f$, thereby approximately equalising their contribution to the total sum. Namely, to sample the missing $z$ coordinate, we determine a smooth non-parametric estimate of the surface density profile of tracers, deproject it to obtain an estimate of the 3d density $\rho(r)$, and sample $z$ from $\rho(\sqrt{R_i^2+z^2})$ at the position of each star $R_i$. For the missing velocity components, we estimate the velocity dispersion of the entire system from the measured values $v_{z,i}$, and sample $v_{x,y}$ from a heavy-tailed bell-like distribution with the same dispersion. The non-uniform sampling is accounted for when computing the contribution $P_i$ of each sample point to the marginalization integral; the estimates of the density profile and velocity dispersion are only used during the initial resampling step, but not in subsequent fitting.

In a spherical system, the integrals of motion are the energy $E$ and angular momentum $L$ per unit mass. Alternatively, one may use actions as the integrals of motion; in this case, $E$ is replaced with the radial action $J_r$, defined as:
\begin{eqnarray}
J_r &\equiv& \frac{1}{\pi}\,\int_{r_{\rm min}}^{r_{\rm max}}\,v_r\,\dd r \nonumber \\
&=& \frac{1}{\pi} \int_{r_\mathrm{min}}^{r_\mathrm{max}} \mathrm{d}r\; \sqrt{2\big[ E - \Phi(r) \big] - L^2/r^2}\ ,
\end{eqnarray}
where $r_\mathrm{min}$ and $r_{\rm max}$ are the roots of the expression under the radical. For each choice of potential, we pre-compute a high-accuracy 2d interpolating spline for efficient evaluation of $J_r(E,L)$, thus using the action variables adds a minor overhead compared to the use of just the classical integrals $E,L$. The advantages of action variables are not apparent in the context of spherical models, but are more important in non-spherical cases.

For the density profile of the DM, responsible for the gravitational potential, we adopt the same five-parameter model in equation \ref{eqn:zhao} that was used to produce the mocks.
On the other hand, we use two different families of DF: one is expressed in terms of $E$ and $L$, and the other -- in terms of radial action $J_r$ and the sum of two other actions $L\equiv J_z+|J_\phi|$; both have six free parameters.

The first DF family is constructed from a given tracer density profile, using the Cuddeford--Osipkov--Merritt inversion formula \citep{1991MNRAS.253..414C}. In this case, we use Equation~\ref{eqn:zhao} for the tracer density profile, which has 4 free parameters: $\alpha_*, \beta_*, \gamma_*$ and $r_*$ (the fifth one, $\nu_0$, is fixed by the normalization constraint). Two other parameters $\beta_0$, $r_a$ define the anisotropy profile (Equation~\ref{eqn:betaprofile} with $n=2$ and $\beta_\infty=1$). This DF family includes the true DF of all variants of models, so we expect to be able to recover the true parameters, given enough data.

The second family is a generalisation of the action-based DF presented in \citet{2015MNRAS.447.3060P}:
\begin{align}   \label{eq:DF_quasiPosti}
f(J_r,L) &= A\,\frac
{\Big(1 + \big[{J_0}/\{p J_r + (1-p)L\}\big]^\eta \Big)^{\Gamma/\eta}\;\;\; }
{\Big(1 + \big[\{q J_r + (1-q)L\}/{J_0}\big]^\eta \Big)^{(\mathrm{B}-\Gamma)/\eta} }\ .
\end{align}
Here the power-law indices $\Gamma$ and B control the steepness of the density profile at small and large radii, correspondingly; $J_0$ determines the transition between the two asymptotic regimes, $\eta$ controls the sharpness of this transition, the mixing parameters $p$ and $q$ are responsible for the velocity anisotropy at small and large radii, and finally $A$ is the overall normalisation constant, which ensures that the integral of $f$ over the entire phase space is unity (it is computed numerically for each choice of six DF parameters and is not a free parameter by itself). 
In this case, there are no separate parameters for the density profile, because it is determined from the DF and the potential: 
$\nu(r) = \iiint \mathrm d^3 \boldsymbol{v}\; 
f\big[ \boldsymbol{I}(\boldsymbol{x}, \boldsymbol{v};\;\Phi) \big]$. However, we do not need the density \textit{per se}, since we fit the joint DF of position and velocity.
Since the true DF used to generate the mock data does not belong to this family of models, we may expect that best-fit parameters of the potential could be biased by the constrained form of the adopted DF.

To explore the parameter space, we use the \EMCEE\ code with 32 walkers evolved for several thousand steps. The evaluation of model likelihoods is performed using \Agama\footnote{The \Agama\ library, together with the complete Python program performing the DF fitting for this study, is available at \url{http://agama.software}}. 

\subsection{Summary of methods and assumptions}

\begin{table*}
\caption{Assumptions and parameters of different methods}
\centering
\tabcolsep 3pt
\begin{tabular}{llllll}
\hline
\hline
Method & {\sc GravSphere} & {\sc DiscreteJAM} & {\sc MAMPOSSt} & \twocol{\sc Agama} \\
\hline
Basis & Binned $v_{\rm LOS}$ $+$ VSPs & Gaussian $v_{\rm LOS}$ & Gaussian $v_{\rm 3D}$ & $f(E,L)$ & $f(J_r,L)$  \\
Radial binning & yes & no & no & \twocol{no} \\
$\rho_{\rm D}(r)$ & 5-piece power law & MGE fit to eq.~(\ref{eqn:zhao}) & eq.~(\ref{eqn:zhao}), $\alpha=1$ & \twocol{eq.~(\ref{eqn:zhao})} \\
$\nu_*(r)$ & sum of Plummer models & MGE fit to eq.~(\ref{eqn:zhao}) & eq.~(\ref{eqn:zhao}), $\alpha=2$, $\beta=5$ & eq.~(\ref{eqn:zhao}) & fitted by DF\\
$\beta_*(r)$ & eq.~(\ref{eqn:betaprofile}) &  eq.~(\ref{eqn:betaprofile}), $n=2$ & eq.~(\ref{eqn:betaprofile}), $n$=2 & eq.~(\ref{eqn:betaprofile}), $n=2, \beta_\infty=1$ & fitted by DF\\
\# of free parameters & 17 & 12 & 9 & 11 & 11 \\
Ranges of free parameters: \\
\qquad $\log_{10} (M_{\rm 200c}/{\rm M}_\odot)$ & unbounded &  unbounded & true$\,\pm\,$1 & \twocol{unbounded} \\
\qquad $r_\DM$/kpc & -- & 0.25 $r_\star$ $\leftrightarrow$  50 $r_\star$ & true$\,\pm\,$0.5 dex & \twocol{$0.1 \leftrightarrow 100$} \\
\qquad $\alpha_{\rm D}$ & -- & $0.5 \leftrightarrow 2.5$ & 1 & \twocol{$0.5 \leftrightarrow 2.5$} \\
\qquad $\beta_{\rm D}$ & -- & $2 \leftrightarrow 5$ & $2 \leftrightarrow 6$ & \twocol{$2.2 \leftrightarrow 6$} \\
\qquad $\gamma_{\rm D}$ & -- & $-0.5 \leftrightarrow \beta_{\rm D}$ & $ -0.1 \leftrightarrow 2$ &  \twocol{$-0.5\leftrightarrow 1.9$} \\
\qquad $\log_{10} (r_\star/\mathrm{kpc})$ & --  & $-2 \leftrightarrow 1$ & true$\,\pm\,$0.3 & $-2 \leftrightarrow 1$ & -- \\
\qquad $\alpha_\star$ & -- & $0.5 \leftrightarrow 2.5$ & 2 & $0.5 \leftrightarrow 3$ & -- \\
\qquad $\beta_\star$ & -- & $2 \leftrightarrow 7$ & 5 & $3.5 \leftrightarrow 6$ & -- \\
\qquad $\gamma_\star$ & -- & $-0.5 \leftrightarrow \beta_\star$ & $-0.05 \leftrightarrow {1.5}$ & $0 \leftrightarrow 1.9$ & -- \\
\qquad $\widetilde \beta_0$ & $-1\leftrightarrow 1$ & $-0.95 \leftrightarrow 1$ & $-0.9 \leftrightarrow 1$ & $-0.5 \leftrightarrow 1$ & -- \\
\qquad $\widetilde \beta_\infty$ & $-1\leftrightarrow 1$ & $-0.95 \leftrightarrow 1$ & $-0.9 \leftrightarrow 1$ & $1$ & --\\
\qquad $\log_{10}(r_a/{\rm kpc})$ & $-2 \leftrightarrow 0$ & ${\log_{10}}({r}_\star) \pm\,$ 0.3& true$\,\pm\,$0.3& $-2 \leftrightarrow 5$ & -- \\
\qquad $n$ & $1 \leftrightarrow 10$ & 2 & 2 & 2 & -- \\
\qquad $\Gamma$ & -- & -- & -- & -- & $0 \leftrightarrow 2.8$ \\
\qquad $\rm B$ & -- & -- & -- & -- & $3.2 \leftrightarrow 12$ \\
Minimization & {\sc emcee} & {\sc emcee} & {\sc CosmoMC} & \twocol{\sc emcee} \\
\# of chains & 1,000 & 100 & 6 & \twocol{32} \\
Length of chains & 5,000 & 20,000 & 90,000 & \twocol{few$\times10^3$} \\
CPU minutes ($10^3$ tracers) & $360$ & $\mathcal O(10^4)^\dagger$ &  $6 \times 15$ (the 6 in $\parallel$) & \twocol{$\mathcal O(10^3)$}\\
\hline
\multicolumn{6}{l}{$^\dagger$The run time speeds up if fewer Gaussians are used; a non-parametric approach would use many fewer Gaussians and could be}\\
\multicolumn{6}{l}{significantly faster.}
\end{tabular}
\label{tab:assump}
\end{table*}

Table~\ref{tab:assump} summarises the different assumptions and parametrisations of the different methods.

\section{Results}\label{sec:results}

In this section, we present results for each method described in \S\ref{sec:methods} for the recovery of the total density and stellar velocity anisotropy at $R = [0.25,0.5,1,2,4]$\,$R_{1/2}$, where $R_{1/2}$ is the projected half light radius, using only line-of-sight velocity data for 1,000 and 10,000 tracers. For brevity, we focus on the `Plum' mocks that are representative of the results from the full suite; for completeness, we present results for the `NonPlum' mocks in Appendix \ref{app:all}. For all mocks, we applied $2\,\rm km\,s^{-1}$ velocity errors to be consistent with current spectroscopic accuracy for dwarf galaxy data (see \S\ref{sec:intro}).

\subsection{Isotropic mocks}

\begin{figure*}
\begin{center}
\includegraphics[width=0.8\textwidth]{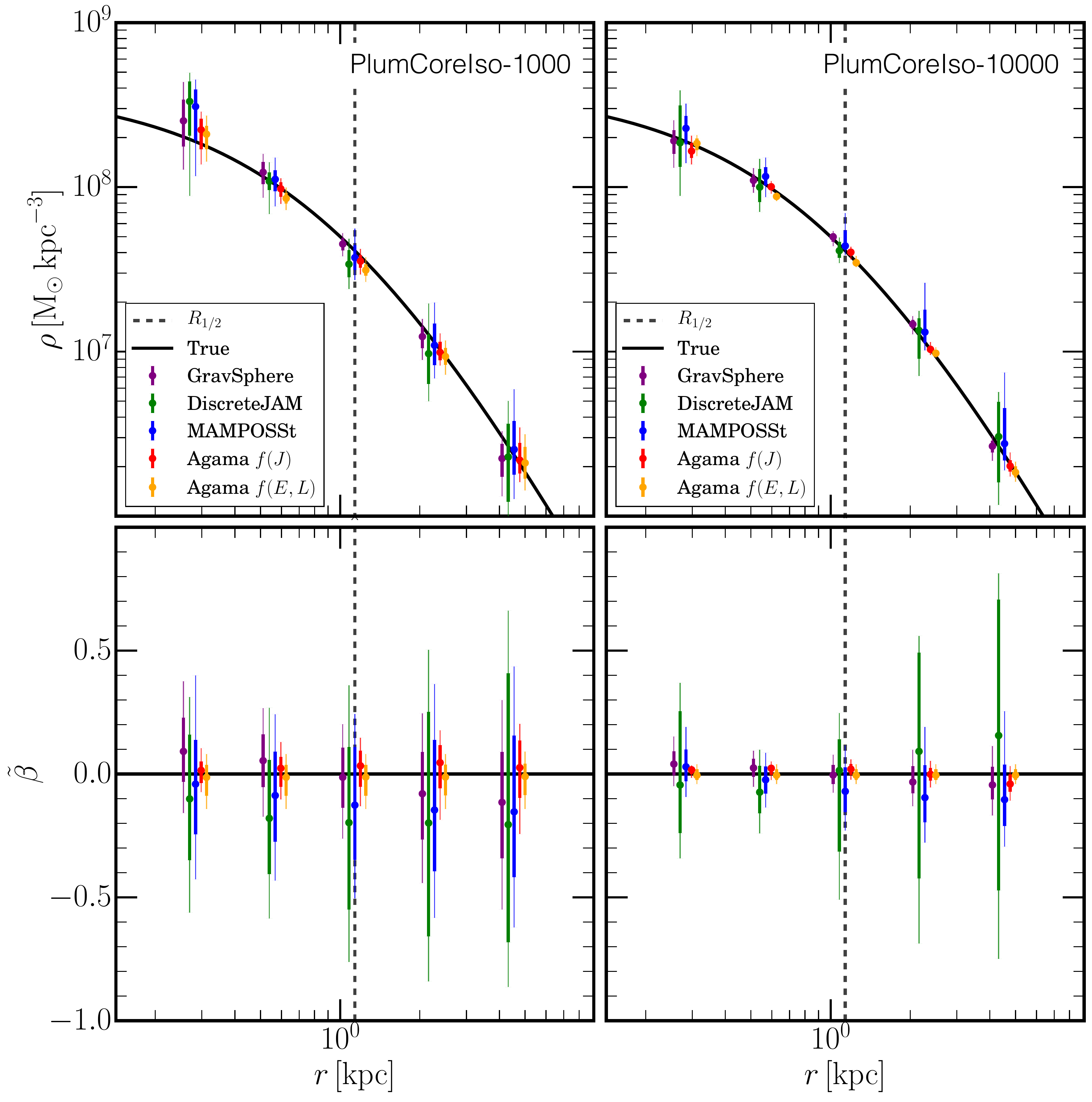}
\caption{Recovery of the density (top) and symmetrised velocity anisotropy (bottom; see equation \ref{eqn:betastar}) for the PlumCoreIso mocks, with line-of-sight velocities for 1,000 (left) and 10,000 (right) tracer stars. The data points show the recovery at $R = [0.25,0.5,1,2,4]$\,$R_{1/2}$, where $R_{1/2}$ is the projected half light radius (vertical dashed line). The thick/thin error bars mark the 68\%/95\% confidence intervals, respectively. The different colours show results for \GravSphere\ (purple), \DiscreteJAM\ (green), \Mamposst\ (blue), \Agama\ $f(\boldsymbol J)$ (red) and $f(E,L)$ (orange), as marked in the legend. Each of the methods is slightly offset left-right from one another to aid clarity (where points are offset, we show the density profile recovery at exactly this offset point). The true mock density and velocity anisotropy profiles are shown by the black lines.}
\label{fig:PlumCoreIso} 
\end{center}
\end{figure*}

\begin{figure*}
\begin{center}
\includegraphics[width=0.8\textwidth]{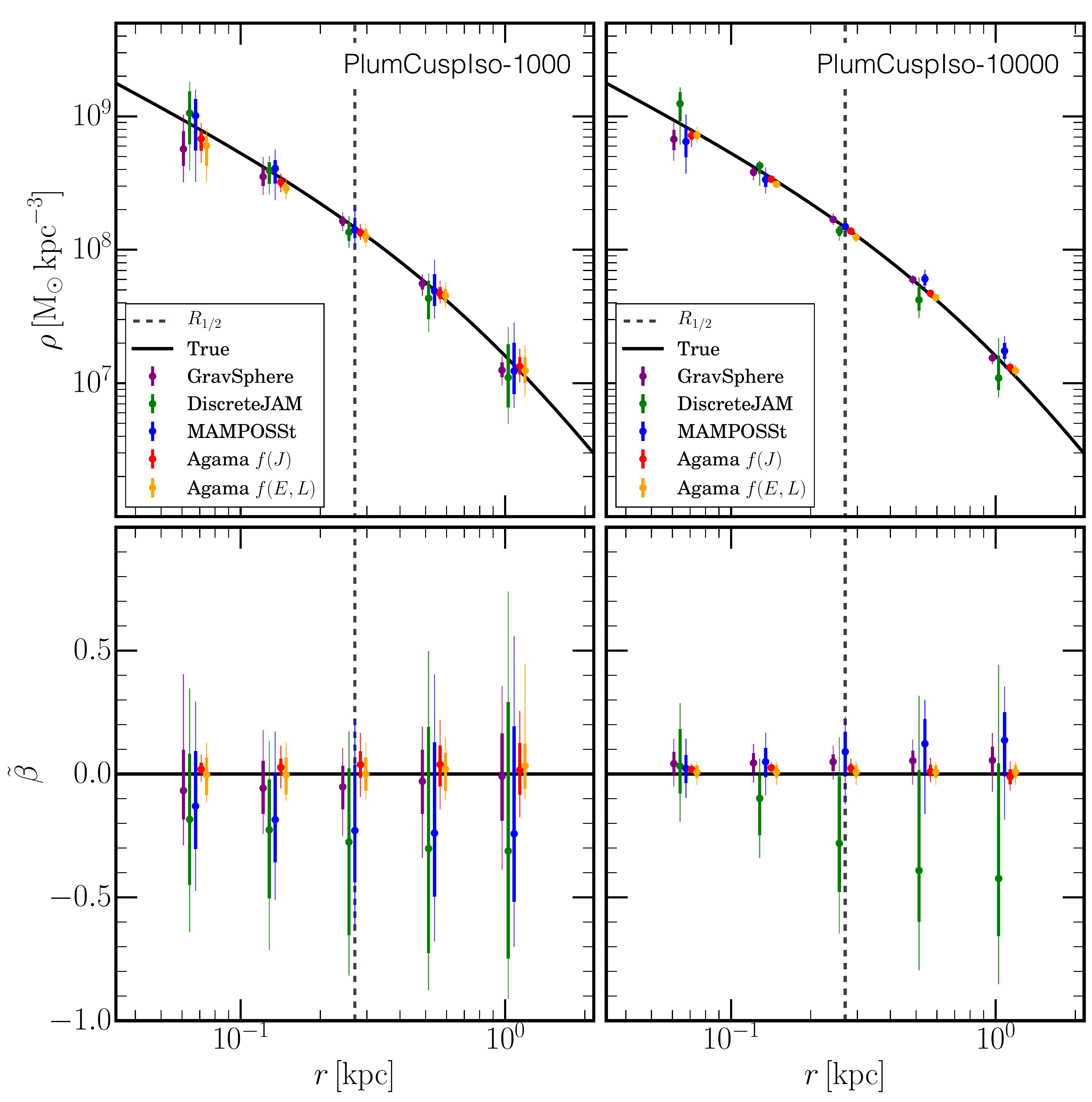}
\caption{As Figure \ref{fig:PlumCoreIso}, but for the PlumCuspIso mock.}
\label{fig:PlumCuspIso} 
\end{center}
\end{figure*}

Figures \ref{fig:PlumCoreIso} and \ref{fig:PlumCuspIso} show the recovery of the density (top) and {\it symmetrised velocity anisotropy} (bottom; see equation \ref{eqn:betastar}) for the isotropic mocks for 1,000 (left) and 10,000 (right) tracers. For these isotropic mocks, all four methods perform similarly well, with the results improving with increased sampling. The density profile is already well-recovered with 1,000 tracers, with all methods recovering the input solution within their 95\% confidence intervals over the radial range $0.25 < R/R_{1/2} < 4$.

However, the uncertainty on $\betastar$ remains substantial for 1,000 tracers. Increasing their number to 10,000, \Agama\ and \GravSphere\ obtain the tightest constraints on $\betastar$, with \Agama\ -- that uses the full shape information in the distribution function -- having the smallest uncertainties. 

\subsection{Radially anisotropic mocks}

\begin{figure*}
\begin{center}
\includegraphics[width=0.8\textwidth]{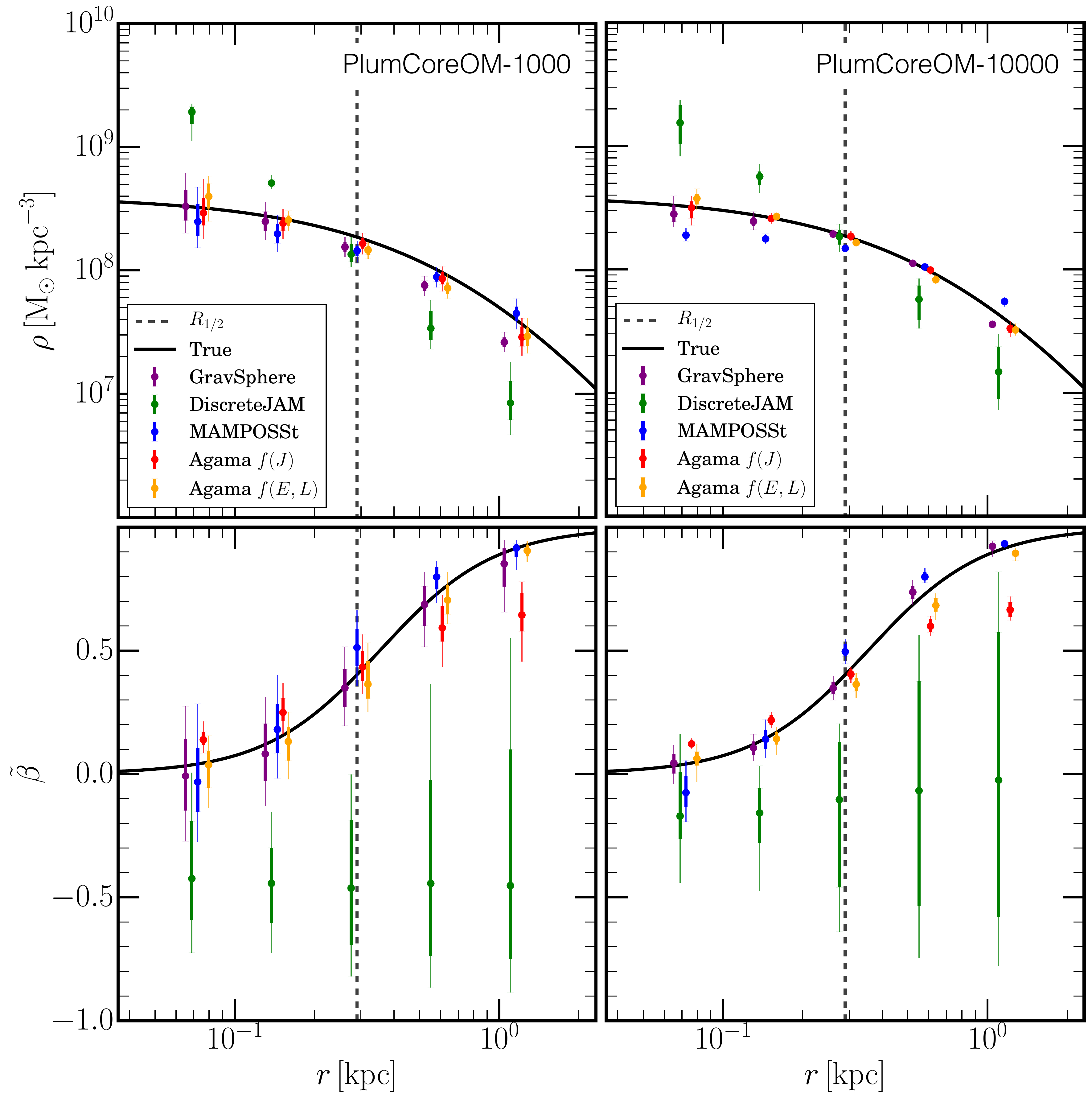}
\caption{As Figure \ref{fig:PlumCoreIso}, but for the PlumCoreOM mock.}
\label{fig:PlumCoreOM}
\end{center}
\end{figure*}

\begin{figure*}
\begin{center}
\includegraphics[width=0.8\textwidth]{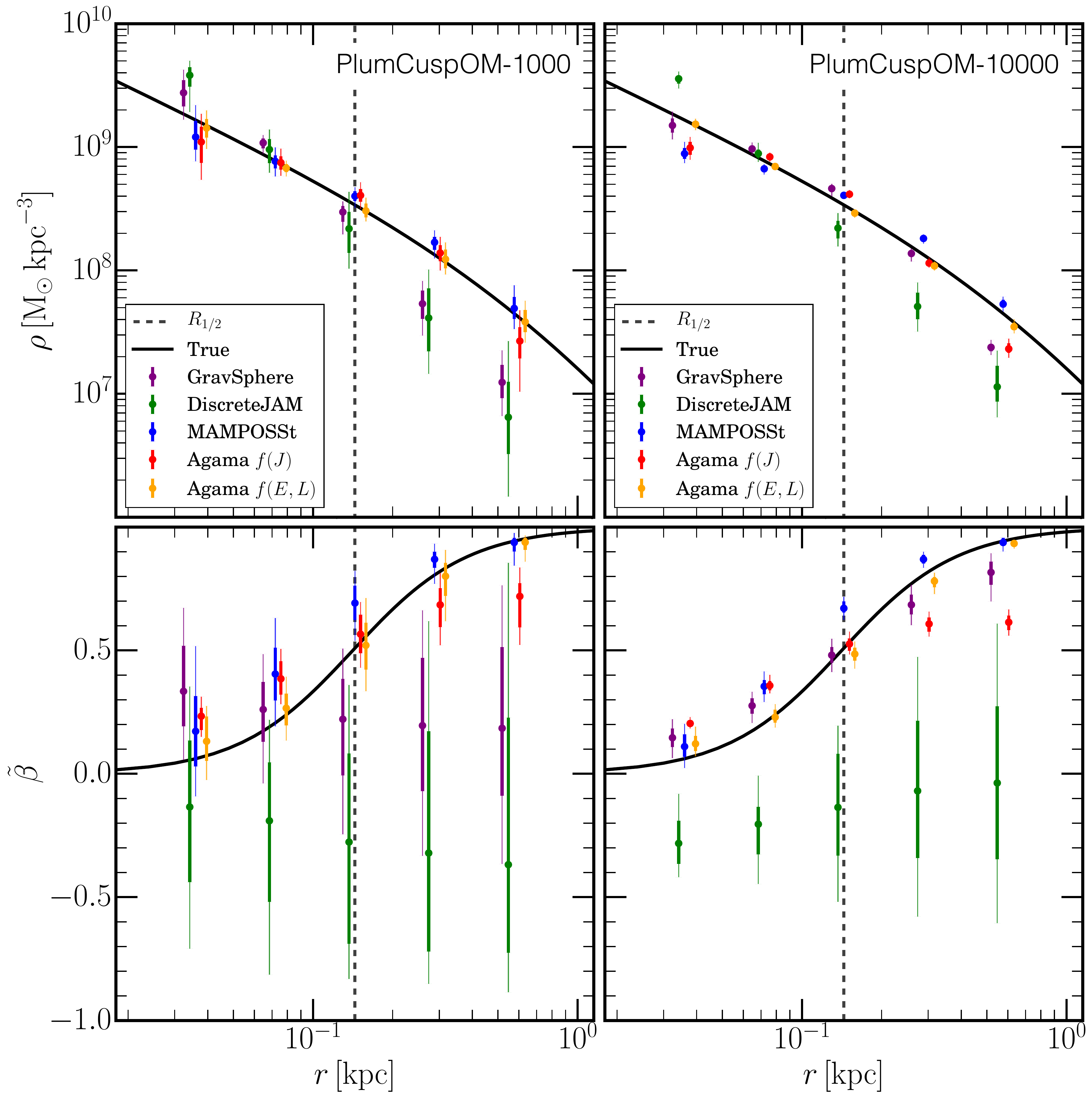}
\caption{As Figure \ref{fig:PlumCoreIso}, but for the PlumCuspOM mock.}
\label{fig:PlumCuspOM} 
\end{center}
\end{figure*}

Moving to radially anisotropic mocks (Figures~\ref{fig:PlumCoreOM} and \ref{fig:PlumCuspOM}), we now see a larger difference between the different methods. Each method recovers some parts of the solution better than others, highlighting the value of modelling the data with multiple techniques. 
Overall, the accuracy of recovery of the potential is tightly linked to the ability of the method to recover the anisotropy profile of the tracers (c.f. \citealt{2017arXiv170104833R}). 

For 1,000 tracers, \Mamposst\ and \Agama\ $f(E,L)$ recover $\betastar(r)$ and $\rho(r)$ over the radial range $0.25 < R/R_{1/2} < 4$ within their 95\% confidence intervals, for both the PlumCoreOM and PlumCuspOM mocks. This may reflect the fact that in these methods, the assumed functional forms for the mass, light and $\betastar$ profiles encompass the true solution (in particular, $\beta_\infty=1$ in the \Agama\ $f(E,L)$ models). By contrast, the \Agama\ $f(\boldsymbol J)$ models do not fully recover the $\betastar$ profile, being more radially anisotropic at small $r$ and less radially anisotropic at large $r$. This negatively impacts the accuracy of potential recovery in the PlumCuspOM case, although not in the PlumCoreOM case. \GravSphere\ recovers the anisotropy profile within its 95\% confidence intervals in the PlumCoreOM case, but performs more poorly in the PlumCuspOM case.
\DiscreteJAM\ and \GravSphere\ are both biased to low $\rho(r)$ beyond $R_{1/2}$, while \DiscreteJAM\ is also biased to high $\rho(r)$ inside $R_{1/2}$ for the PlumCoreOM mock. This owes to a bias towards tangential anisotropy for the \DiscreteJAM\ method in both PlumCoreOM and PlumCuspOM mocks that does not diminish even for 10,000 tracer stars. This behaviour is also seen in the \GravSphere\ code if the VSPs are not used in the fit. \citet{2017arXiv170104833R} show that this bias owes to the mass-anisotropy degeneracy, combined with the true solution for this mock lying on the edge of the hypervolume of acceptable models. Indeed, they show that the correct model {\it is} recovered within the full MCMC chains, but is rare as compared to the large number of tangentially anisotropic models that fit the data similarly well.

Finally, moving to 10,000 tracers, bias starts to creep in or become more statistically significant in most of the methods. The \Agama\ $f(E,L)$ models still produce good fits within their 95\% confidence intervals, while the $f(\boldsymbol J)$ models retain the biases in the anisotropy profile in both cases, and in $\rho(r)$ in the PlumCuspOM case, reflecting the fact that the assumed distribution function differs from the one used to set up the mocks (c.f. the earlier discussion on this in \S\ref{sec:intro}). \Mamposst\ also starts to show similarly biased behaviour, with the inner density underestimated at small radii for the PlumCuspOM and PlumCoreOM mocks, and bias now present in the recovery of $\betastar(r)$. This likely owes to the assumed Gaussian form of the local distribution function being a good, but not perfect, approximation to its true shape. The \DiscreteJAM\ method -- that is not able to break the mass-anisotropy degeneracy with line-of-sight data alone -- retains the bias seen in its recovery of the 1,000 star mocks, but with smaller error bars. The \GravSphere\ method -- that is designed to make as few assumptions as possibly about the form of the gravitational potential and the distribution function -- approaches the true solution more closely, although some bias towards lower $\rho(r)$ remains at large radii for the PlumCuspOM mock. This bias is, however, significantly reduced as compared to the 1,000 star solution.

Finally, we note that some of the differences in performance between the different methods owes also to their different priors (see Table \ref{tab:assump}). \GravSphere\ has the loosest priors of all methods presented here and, correspondingly, the greatest challenge in recovering the mocks. By contrast, \Mamposst\ assumes tighter priors that are symmetrised around the true solution, \DiscreteJAM\ uses priors for the anisotropy and DM scale radii that are tightly linked to the stellar density scale radius (which is itself very well recovered), and \Agama\ $f(E,L)$ assumes $\beta_\infty = 1$, as in the mocks. Moreover, \DiscreteJAM, \Mamposst, and \Agama\ each assume the same functional form for the mass profile as in the mock data (with \Mamposst\ fixing the value of $\alpha_D$), whereas \GravSphere\ does not. As such, we cannot really speak of any one method being `superior' to the others; rather, each has its strengths and weaknesses.

\subsection{Tangentially anisotropic mocks}

\begin{figure*}
\begin{center}
\includegraphics[width=0.8\textwidth]{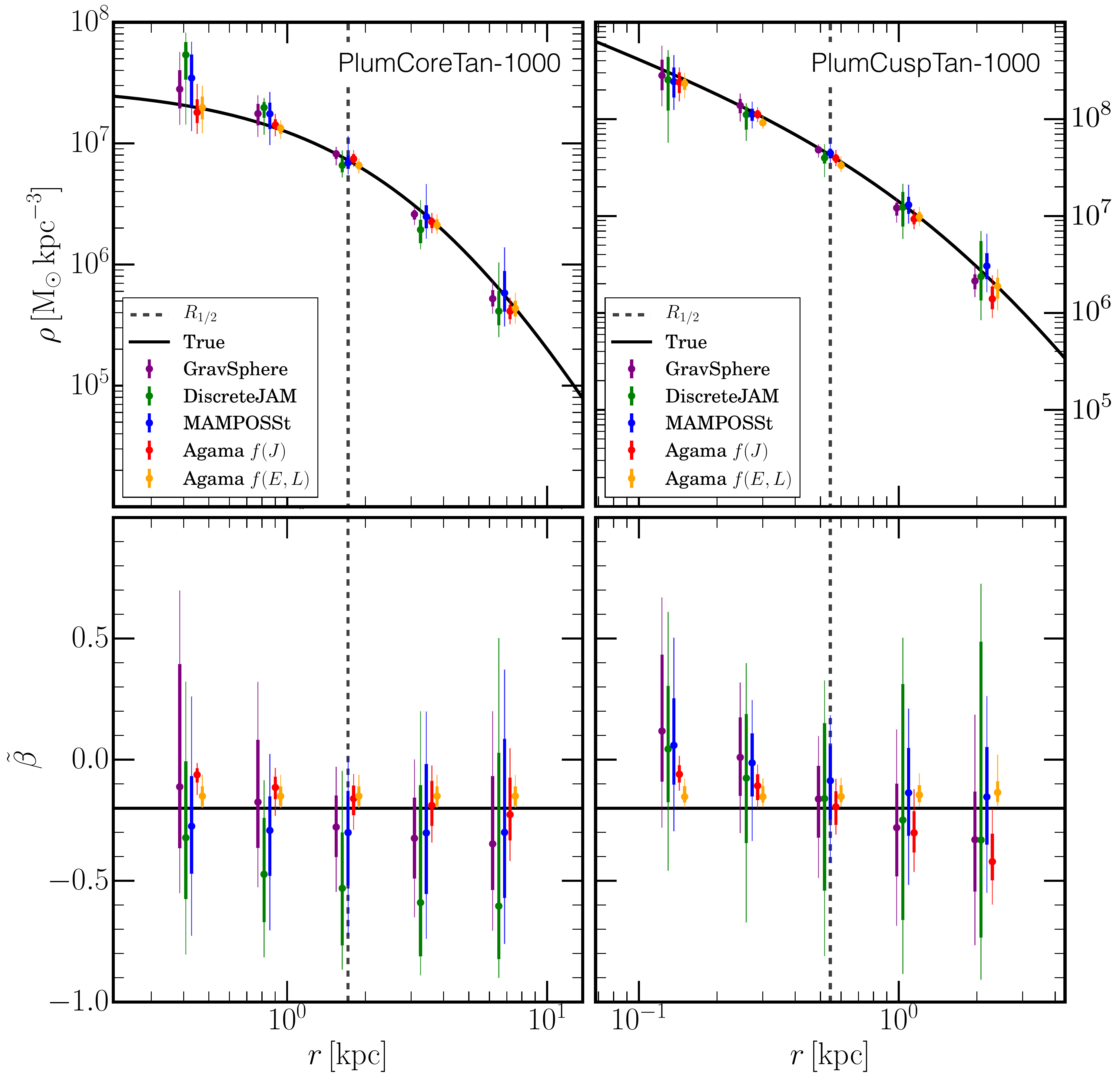}
\caption{As Figure \ref{fig:PlumCoreIso}, but for mocks PlumCoreTan (left) and PlumCuspTan (right) with 1000 tracer stars.}
\label{fig:composite_tan} 
\end{center}
\end{figure*}

In Figure \ref{fig:composite_tan}, we show our results for the tangentially anisotropic mocks, PlumCoreTan and PlumCuspTan assuming 1000 tracer stars. Notice that, similarly to the isotropic mocks, all methods recover $\rho(r)$ within their 95\% confidence intervals over the range $0.25 < R/R_{1/2} < 4$ (top panels), though for the PlumCoreTan mock, \DiscreteJAM\ is biased (at 68\% confidence) towards cuspy, tangentially anisotropic mocks and \Mamposst\ marginally so. This is similar, though less severe, to the bias seen for the PlumCoreOm mock and owes to the mass-anisotropy degeneracy (see discussion above).

All methods successfully detect the tangential anisotropy, though for 1,000 tracers, this is only statistically significant near the half-light radius (vertical dashed line). Interestingly, both the \Agama\ $f(E,L)$ and the \Agama\ $f(\boldsymbol J)$ methods become biased at greater than their 95\% confidence intervals. \Agama\ $f(\boldsymbol J)$ is overly isotropic in the inner regions and, for PlumCuspTan, overly tangential in its outer regions, while \Agama\ $f(E,L)$ is overly isotropic at all radii. These biases reflect the assumed parameterisation of the distribution function and the choice of priors. As discussed above and in \S\ref{sec:intro}, this will lead to bias if the phase space distribution function of the mock data is inconsistent with these assumptions, as is the case here. Indeed, the tangentially anisotropic mocks present a particular challenge for the \Agama\ $f(E,L)$ models, since the true value of $\beta=-0.5$ lies at the boundary of the allowed range (for technical reasons, anisotropic DFs computed by the Cuddeford--Osipkov--Merritt method are restricted to have $\widetilde\beta_0\ge -0.5$).

\subsection{Adding proper motion data}\label{sec:propermotions}

\begin{figure*}
\begin{center}
\includegraphics[width=0.8\textwidth]{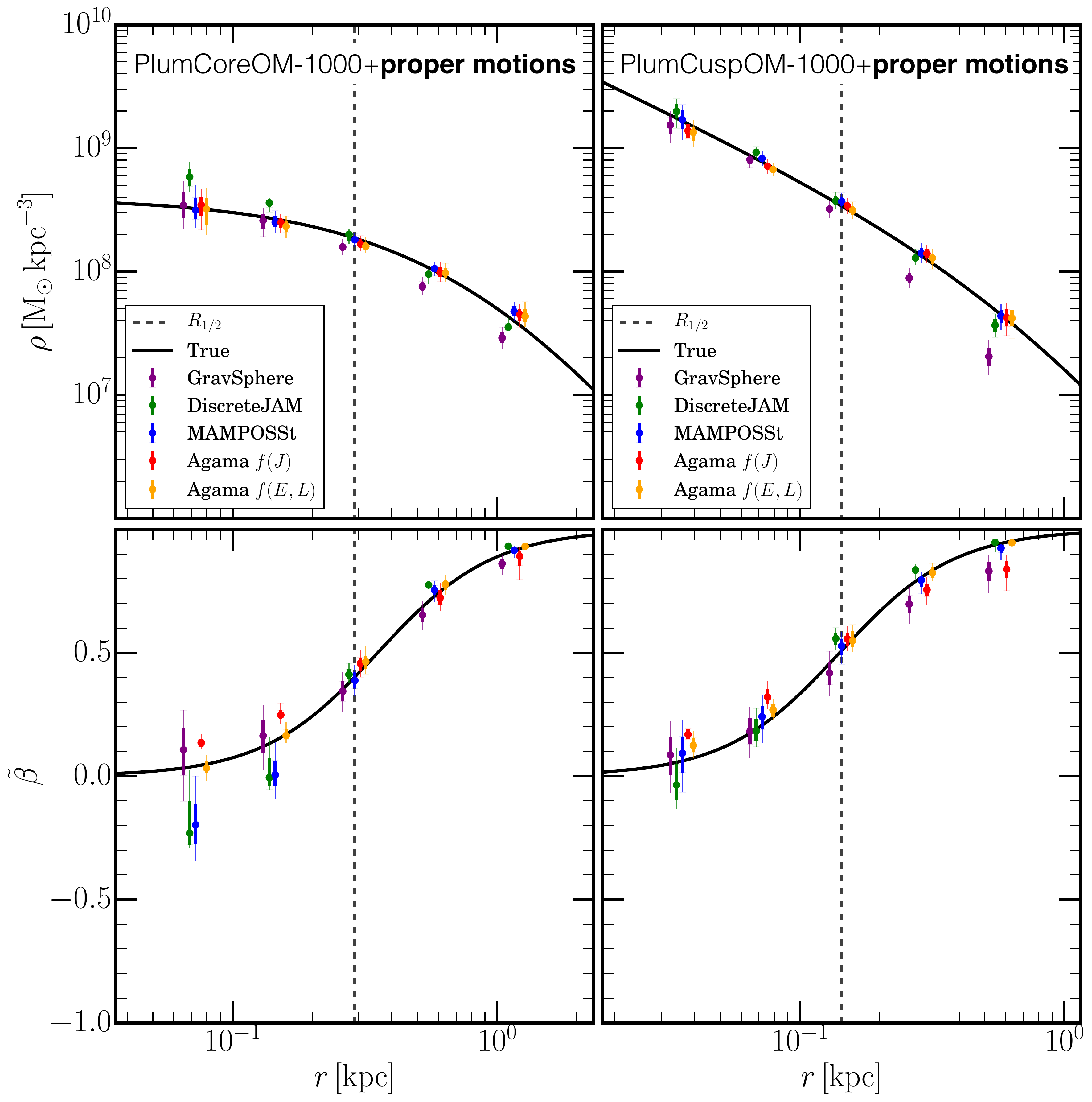}
\caption{As Figure \ref{fig:PlumCoreIso}, but for mocks PlumCoreOM (left) and PlumCuspOM (right) including also proper motion data (see text for details).}
\label{fig:composite_prop} 
\end{center}
\end{figure*}

In Figure \ref{fig:composite_prop}, we show the performance of all methods when adding proper motion data. For this, we consider 1,000 tracers for the PlumCuspOM and PlumCoreOM mocks assuming 2\,km\,s$^{-1}$ errors on all three components of the velocity (the other mock data produced comparable results). All three Jeans models constrain the line of sight, radial and tangential dispersion profiles as in \citet{2017arXiv170104833R}, while the \Agama\ DF method uses the 5D phase-space data. As can be seen in Figure \ref{fig:composite_prop}, most methods now recover both $\rho(r)$ and $\beta(r)$ within their quoted 95\% confidence intervals, although some biases still remain. \GravSphere\ underestimates $\rho(r)$ at large radii, while \DiscreteJAM\ overestimates $\rho(r)$ at small radii for the PlumCoreOM mock. \Agama\ $f(\boldsymbol J)$ and \GravSphere\ produce less radially anisotropic $\betastar$ profiles at large radii for the PlumCuspOM mock, and \Agama\ $f(\boldsymbol J)$ models are slightly radially biased in the centre. \Mamposst\ obtains a good recovery of $\rho(r)$ and the correct shape for $\betastar(r)$, but has $\betastar(r)$ biased low at small radii in PlumCuspOM. Nonetheless, with data of this quality all methods are able to distinguish the cusped and cored mocks at high confidence.

\section{Discussion \& Conclusions}\label{sec:conclusions}

\begin{figure*}
\begin{center}
\includegraphics{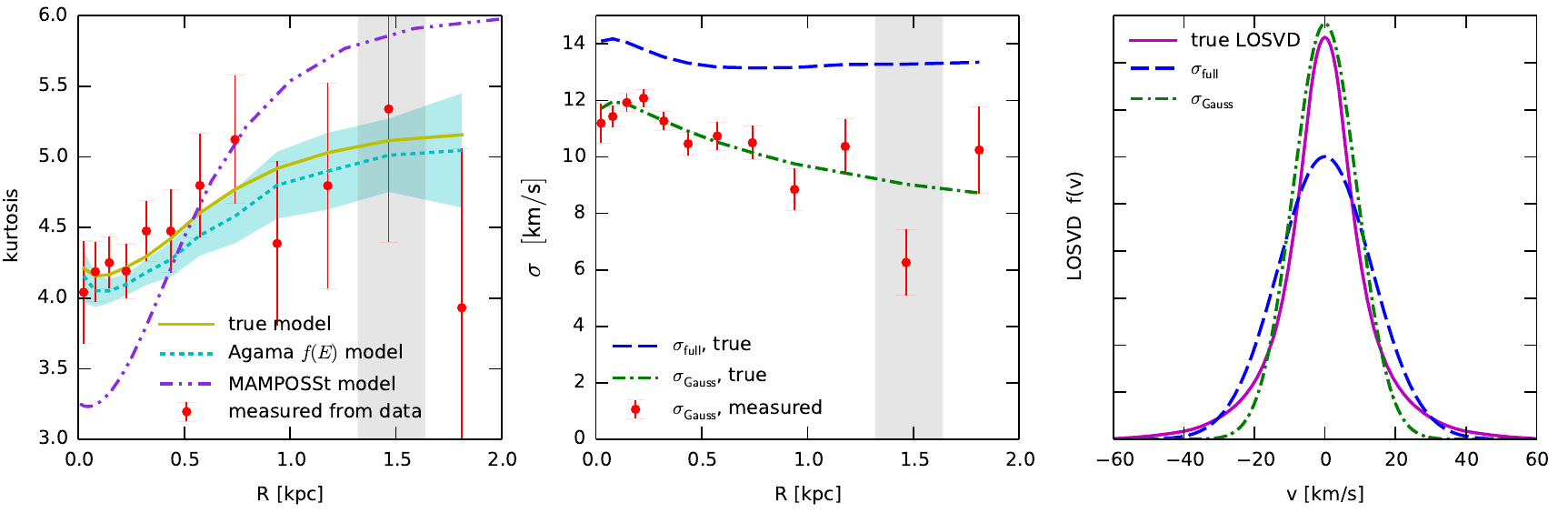}
\caption{Illustration of non-Gaussianity of LOSVD in a radially-anisotropic model (PlumCoreOM).  \protect\\   
\textit{Left panel:} radial profile of the line-of-sight kurtosis $\kappa \equiv \overline{v^4} / \sigma_\mathrm{full}^4$, where $\sigma_\mathrm{full}^2 \equiv \overline{v^2}$ is the line-of-sight velocity dispersion (full second moment of the velocity distribution). For a Gaussian LOSVD, $\kappa=3$; a higher value indicates a fat-tailed distribution. Solid yellow line is the true $\kappa$ in the given model; red points with error bars are the binned measured values in the $10^4$-star mock dataset; dotted cyan line with a shaded region is the mean and the $68\%$ interval of values in the \Agama\ $f(E,L)$ models; dash-double-dotted purple line is the \Mamposst\ model prediction (neglecting velocity errors). It is clear that the LOSVD is significantly non-Gaussian with fat tails across the entire range of radii. The DF-based models are able to capture this behaviour, while the \Mamposst\ method reproduces the trend qualitatively but not in detail, underlining the limitations of using the Gaussian assumption even for the local (not projected) velocity distribution. Gray-shaded region indicates the range of radii used for the LOSVD in the rightmost panel. \protect\\
\textit{Centre panel:} radial profile of the full line-of-sight velocity dispersion $\sigma_\mathrm{full}$ (dashed blue line) and the width of the best-fit Gaussian $\sigma_\mathrm{Gauss}$ (dot-dashed green line). The difference between the two curves indicates the non-Gaussian shape of the LOSVD: the width of the best-fit Gaussian is comparable to the width of the main peak of the LOSVD, but is significantly smaller than the true second moment of the LOSVD, which is heavily influenced by its tails. Red points with error bars are the binned measured values of the Gaussian approximation to the LOSVD in the $10^4$-star mock dataset. \protect\\
\textit{Right panel:} LOSVD in the penultimate bin (the range of radii indicated by the shaded gray region in the other panels). Solid magenta curve is the true LOSVD; dashed blue curve is a Gaussian profile with the width given by the full second moment of the LOSVD $\sigma_\mathrm{full}$ (the quantity that enters the Jeans equations); dot-dashed green curve is the best-fit Gaussian with width $\sigma_\mathrm{Gauss}$. Neither is a good approximation to the actual LOSVD, which has prominent fat tails.}
\label{fig:LOSVD} 
\end{center}
\end{figure*}

We have shown that all four mass modelling methods considered in this paper, using just line-of-sight velocity data, are able to recover both the density and velocity anisotropy as a function of radius  within their 95\% confidence intervals over the radial range $0.25 < R/R_{1/2} < 4$, for spherically symmetric mock stellar systems, provided that the mocks are isotropic or tangentially anisotropic. However, strong radial anisotropy at large radii presents a more challenging test. Only methods that utilise some information about the shape of the distribution function are able to recover the density and velocity anisotropy profiles of these mocks. \GravSphere\ achieves this by using the two fourth order `Virial Shape Parameters' \citep{1990AJ.....99.1548M}; \Agama\ achieves it by directly fitting a distribution function; and \Mamposst\ achieves it by projecting an assumed local Gaussian velocity distribution function along the line of sight. To illustrate this, in Figure \ref{fig:LOSVD} we show the radial profiles of line-of-sight velocity dispersion and kurtosis $\kappa$, as well as the line-of-sight velocity distribution (LOSVD) in the outer parts, for the PlumCoreOM model. Notice that the profiles are substantially non-Gaussian (fat-tailed) across the entire range of radii, indicated by an elevated value of $\kappa$. The DF-based \Agama\ models are able to capture this behaviour, while the projected Jeans equations used in the \DiscreteJAM\ method, which uses only the second moment of the distribution function, assuming a Gaussian LOSVD, become biased.
Also, \Mamposst\ predicts too low and high values of kurtosis at small and large radii, respectively, explaining its respective underestimate and overestimate of $\beta$ at small and large radii (right panels of Fig.~\ref{fig:PlumCoreOM}).

Adding internal proper motion data for 1,000 tracer stars, most methods recovered both $\rho(r)$ and $\beta(r)$ within their 95\% confidence intervals over the radial range $0.25 < R/R_{1/2} < 4$, albeit with some small biases at the largest and smallest radii. (\GravSphere, in particular, underestimated the density for $R > 2 R_{1/2}$ on these tests.) These findings are in good agreement with previous work \citep[e.g.][]{2007ApJ...657L...1S, 2017arXiv170104833R}, demonstrating the value of obtaining proper motion data for significant numbers of individual stars in nearby dwarf galaxies. This is just beginning to become possible now with the advent of combined \textit{Hubble Space Telescope} and \textit{Gaia} proper motions \citep{2018NatAs...2..156M, 2020A&A...633A..36M,Vitral&Mamon20}, but really large datasets will have to wait for the upcoming \textit{Nancy Grace Roman Space Telescope} \citep{WFIRST2017} and proposed space astrometry missions like Theia \citep{2017arXiv170701348T}.

We conclude that breaking the mass-anisotropy degeneracy is crucial for obtaining an unbiased measurement of the density profile and velocity anisotropy of spherical stellar systems. Previous work has focused on achieving this using multiple tracer populations with different scale lengths \citep[e.g.][]{2008ApJ...681L..13B, 2011ApJ...742...20W, 2012MNRAS.419..184A, 2017arXiv170104833R}. While this has been shown to work well, it can only be applied to systems that have distinct populations. In this paper, we have focused instead on modelling single tracer population data. In this case, with only line-of-sight velocity data, we have shown that the degeneracy can be broken by using higher order `Virial Shape Parameters', by directly fitting a global distribution function, or by self-consistently projecting an assumed local velocity distribution function along the line of sight. We have also uncovered interesting differences even between rather similar-looking methodologies, highlighting the utility of performing detailed tests on mock data and exploring different mass modelling approaches, as we have done here.

Finally, we note that all of the tests in this paper assumed perfect data with small (2\,km/s) velocity errors. Already this was challenging for some methods. Real stellar systems will have additional uncertainties from binary star and foreground contamination \citep[e.g.][]{2011ApJ...742...20W, 2018MNRAS.481..860R}, disequilibrium due to tides \citep[e.g.][]{2015NatCo...6E7599U, 2013MNRAS.431.2796K, 2018MNRAS.481..860R, 2019arXiv191109124G},  rotation \citep[e.g.][]{Watkins2013, 2016MNRAS.463.1117Z}, departures from spherical symmetry \citep[e.g.][]{Mamon+13,2013MNRAS.431.2796K, 2017arXiv170104833R, 2019arXiv191109124G, 2020arXiv200713780H}, and inconsistency between the tracer density and kinematic measurements \citep[e.g.][]{2014JPhG...41f3101R}. Some of these problems can be ameliorated with better data, for example, repeat measurements of individual stars to explore binary contamination \citep[e.g.][]{2011ApJ...736..146K, 2018AJ....156..257S}, metallicity measurements to improve foreground separation, and having a sufficient number of unbiased stars with kinematics that the photometric and kinematic samples are identical. However, some problems like tides will remain, even with exquisite data \citep[e.g.][]{2013MNRAS.431.2796K, 2015NatCo...6E7599U}. We will explore these issues further in future work.

\section{Acknowledgements}

We would like to thank the anonymous referee for a constructive and thorough report that improved the paper. JIR would like to acknowledge support from STFC consolidated grant ST/M000990/1 and the MERAC foundation. GAM thanks the French Programme National de Cosmologie et Galaxies (PNCG) of CNRS/INSU with INP and IN2P3, co-funded by CEA and CNES, for financial support of a workshop at IAP during which this work was finalized. EV and PD acknowledge support from the European Research Council (ERC) Horizon 2020 programme under grants 321067 (EV and PD) and 308024 (EV). LLW acknowledges support provided by NASA through \textit{HST} grants AR-14322 and AR-15055, from the Space Telescope Science Institute, which is operated by the Association of Universities for Research in Astronomy, Inc., under NASA contract NAS 5-26555, and from the ERC Horizon 2020 programme under grant agreement No 724857 (Consolidator Grant ArcheoDyn).

This research has made use of Astropy\footnote{\url{http://www.astropy.org}}, a community-developed core Python package for Astronomy \citep{astropy2013},  \PyNbody\footnote{\url{https://github.com/pynbody/pynbody}} for the simulation analysis \citep{2013ascl.soft05002P},  \EMCEE\footnote{\url{https://github.com/dfm/emcee}} \citep{2013PASP..125..306F}, and CosmoMC\footnote{\url{https://cosmologist.info/cosmomc/}} 
\citep{Lewis&Bridle02} for MCMC sampling, as well as NASA's Astrophysics Data System.

\appendix

\section{Results for the remainder of the spherical mocks}\label{app:all}

\begin{figure*}
\begin{center}
\includegraphics[width=0.99\textwidth]{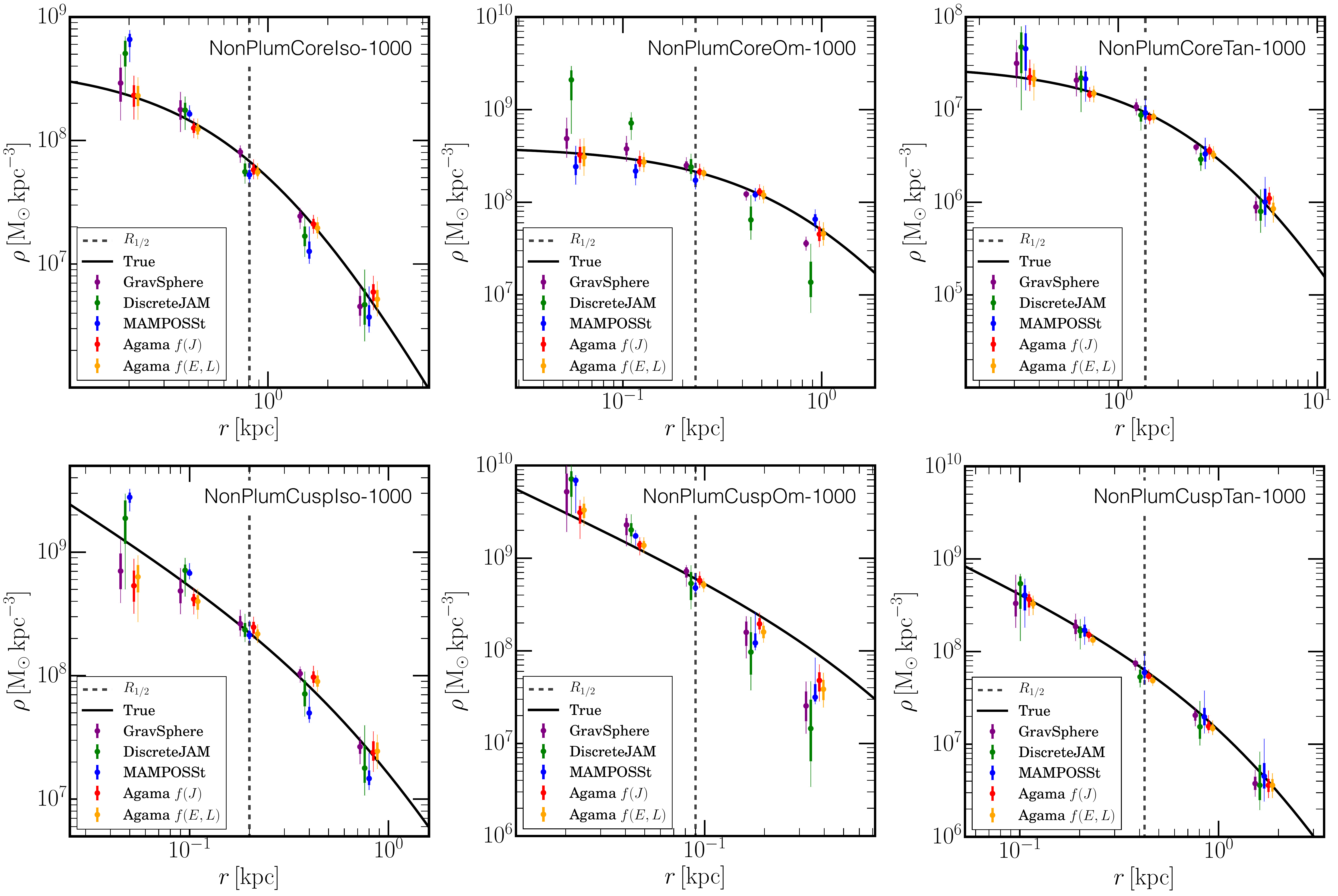}
\caption{As Figure \ref{fig:PlumCoreIso} but for the `NonPlum' spherical mocks, as marked on the panels.}
\label{fig:composite_all} 
\end{center}
\end{figure*}

For completeness, in this Appendix we show the spherically averaged density profile for the remainder of the spherical mocks with 1000 tracer stars, the `NonPlum' models (Figure \ref{fig:composite_all}).

As can be seen, the results are broadly consistent with those for the similar `Plum' mocks, with \DiscreteJAM\ showing a bias towards cuspy, tangentially anisotropic models, particularly for NonPlumCoreOm, while \Mamposst\ shows an even stronger cuspy bias for NonPlumCoreIso. As discussed in \S\ref{sec:results} and \S\ref{sec:conclusions}, this owes to the mass-anisotropy degeneracy.

Similarly to the PlumCuspOm mock, \GravSphere\ and \DiscreteJAM\ fall off too steeply at large radii for NonPlumCuspOm, however \Mamposst\ and the \Agama\ models also show a similar, albeit less severe, bias. 

Interestingly, all of the models except \DiscreteJAM\ show a slight bias towards cores in the innermost region ($0.25\,R_{1/2}$) for the NonPlumCuspIso mock. \DiscreteJAM\ does not show this bias, but this is because it shows a bias instead towards tangentially anisotropic models.
(Tangentially anisotropic models require a steeper density profile to match the data that, in this case, drives a better match with the input model.) 
\Mamposst\ (and \DiscreteJAM\ to a lesser extent) finds instead an even steeper inner density profile than the NFW model used to build this mock. By contrast, for the NonPlumCuspIso mock, both \GravSphere\ and \Agama\ are consistent with the input isotropic distribution function within their 95\% confidence intervals, while \Mamposst\ produces tangential orbits at all radii (not shown for brevity). As such, the bias towards cores in the NonPlumCuspIso mock does not owe to the mass-anisotropy degeneracy. We speculate that it owes instead to the cuspy light profile in this mock yielding a poorer radial sampling of the stellar kinematics. (For a fixed number of tracers, cuspier light profiles will have relatively fewer sample points at large radii.) This warrants further investigation in future work. However, we note that it is unlikely to impact most real stellar systems, since these typically have a much shallower inner tracer density profile than the rather extreme `NonPlum' mocks explored here \citep[see e.g.][for nearby dwarf galaxies]{2019MNRAS.484.1401R}.

\bibliographystyle{mn2e}
\bibliography{refs}

\begin{thebibliography}{82}
\expandafter\ifx\csname natexlab\endcsname\relax\def\natexlab#1{#1}\fi

\bibitem[{{Amorisco} \& {Evans}(2012)}]{2012MNRAS.419..184A}
{Amorisco} N.~C., {Evans} N.~W., 2012, \mnras, 419, 184

\bibitem[{{An} \& {Evans}(2006)}]{2006ApJ...642..752A}
{An} J.~H., {Evans} N.~W., 2006, \apj, 642, 752

\bibitem[{{Astropy Collaboration} {et~al}\mbox{.}(2013){Astropy Collaboration},
  {Robitaille}, {Tollerud}, {Greenfield}, {Droettboom}, {Bray}, {Aldcroft},
  {Davis}, {Ginsburg}, {Price-Whelan}, {Kerzendorf}, {Conley}, {Crighton},
  {Barbary}, {Muna}, {Ferguson}, {Grollier}, {Parikh}, {Nair}, {Unther},
  {Deil}, {Woillez}, {Conseil}, {Kramer}, {Turner}, {Singer}, {Fox}, {Weaver},
  {Zabalza}, {Edwards}, {Azalee Bostroem}, {Burke}, {Casey}, {Crawford},
  {Dencheva}, {Ely}, {Jenness}, {Labrie}, {Lim}, {Pierfederici}, {Pontzen},
  {Ptak}, {Refsdal}, {Servillat}, \& {Streicher}}]{astropy2013}
{Astropy Collaboration} {et~al.}, 2013, \aap, 558, A33

\bibitem[{{Battaglia} {et~al}\mbox{.}(2008){Battaglia}, {Helmi}, {Tolstoy},
  {Irwin}, {Hill}, \& {Jablonka}}]{2008ApJ...681L..13B}
{Battaglia} G., {Helmi} A., {Tolstoy} E., {Irwin} M., {Hill} V., {Jablonka} P.,
  2008, \apjl, 681, L13

\bibitem[{{Bellini} {et~al}\mbox{.}(2014){Bellini}, {Anderson}, {van der
  Marel}, {Watkins}, {King}, {Bianchini}, {Chanam{\'e}}, {Chandar}, {Cool},
  {Ferraro}, {Ford}, \& {Massari}}]{Bellini2014}
{Bellini} A. {et~al.}, 2014, \apj, 797, 115

\bibitem[{{Binney} \& {Mamon}(1982)}]{Binney&Mamon82}
{Binney} J., {Mamon} G.~A., 1982, \mnras, 200, 361

\bibitem[{{Binney} \& {Tremaine}(2008)}]{2008gady.book.....B}
{Binney} J., {Tremaine} S., 2008, {Galactic Dynamics: Second Edition}.
  Princeton Univ. Press

\bibitem[{{Bonnivard} {et~al}\mbox{.}(2015){Bonnivard}, {Combet}, {Daniel},
  {Funk}, {Geringer-Sameth}, {Hinton}, {Maurin}, {Read}, {Sarkar}, {Walker}, \&
  {Wilkinson}}]{2015MNRAS.453..849B}
{Bonnivard} V. {et~al.}, 2015, \mnras, 453, 849

\bibitem[{{Cappellari}(2002)}]{Cappellari2002}
{Cappellari} M., 2002, \mnras, 333, 400

\bibitem[{{Cappellari}(2008)}]{Cappellari2008}
{Cappellari} M., 2008, \mnras, 390, 71

\bibitem[{{Cappellari}(2015)}]{Cappellari2015}
{Cappellari} M., 2015, ArXiv:1504.05533

\bibitem[{{Cappellari} {et~al}\mbox{.}(2012){Cappellari}, {McDermid},
  {Alatalo}, {Blitz}, {Bois}, {Bournaud}, {Bureau}, {Crocker}, {Davies},
  {Davis}, {de Zeeuw}, {Duc}, {Emsellem}, {Khochfar}, {Krajnovi{\'c}},
  {Kuntschner}, {Lablanche}, {Morganti}, {Naab}, {Oosterloo}, {Sarzi}, {Scott},
  {Serra}, {Weijmans}, \& {Young}}]{2012Natur.484..485C}
{Cappellari} M. {et~al.}, 2012, \nat, 484, 485

\bibitem[{{Chiappo} {et~al}\mbox{.}(2017){Chiappo}, {Cohen-Tanugi}, {Conrad},
  {Strigari}, {Anderson}, \& {S{\'a}nchez-Conde}}]{2017MNRAS.466..669C}
{Chiappo} A., {Cohen-Tanugi} J., {Conrad} J., {Strigari} L.~E., {Anderson} B.,
  {S{\'a}nchez-Conde} M.~A., 2017, \mnras, 466, 669

\bibitem[{{Cuddeford}(1991)}]{1991MNRAS.253..414C}
{Cuddeford} P., 1991, \mnras, 253, 414

\bibitem[{{Dehnen}(2009)}]{2009MNRAS.395.1079D}
{Dehnen} W., 2009, \mnras, 395, 1079

\bibitem[{{Diakogiannis} {et~al}\mbox{.}(2019){Diakogiannis}, {Lewis}, {Ibata},
  {Guglielmo}, {Wilkinson}, \& {Power}}]{2019MNRAS.482.3356D}
{Diakogiannis} F.~I., {Lewis} G.~F., {Ibata} R.~A., {Guglielmo} M., {Wilkinson}
  M.~I., {Power} C., 2019, \mnras, 482, 3356

\bibitem[{{Eggen} {et~al}\mbox{.}(1962){Eggen}, {Lynden-Bell}, \&
  {Sandage}}]{1962ApJ...136..748E}
{Eggen} O.~J., {Lynden-Bell} D., {Sandage} A.~R., 1962, \apj, 136, 748

\bibitem[{{Emsellem} {et~al}\mbox{.}(1994){Emsellem}, {Monnet}, \&
  {Bacon}}]{Emsellem1994}
{Emsellem} E., {Monnet} G., {Bacon} R., 1994, \aap, 285, 723

\bibitem[{{Foreman-Mackey} {et~al}\mbox{.}(2013){Foreman-Mackey}, {Hogg},
  {Lang}, \& {Goodman}}]{2013PASP..125..306F}
{Foreman-Mackey} D., {Hogg} D.~W., {Lang} D., {Goodman} J., 2013, \pasp, 125,
  306

\bibitem[{{Garbari} {et~al}\mbox{.}(2011){Garbari}, {Read}, \&
  {Lake}}]{2011MNRAS.416.2318G}
{Garbari} S., {Read} J.~I., {Lake} G., 2011, \mnras, 416, 2318

\bibitem[{{Genina} \& {Fairbairn}(2016)}]{2016MNRAS.463.3630G}
{Genina} A., {Fairbairn} M., 2016, \mnras, 463, 3630

\bibitem[{{Genina} {et~al}\mbox{.}(2020){Genina}, {Read}, {Frenk}, {Cole},
  {Ben{\'\i}tez-Llambay}, {Ludlow}, {Navarro}, {Oman}, \&
  {Robertson}}]{2019arXiv191109124G}
{Genina} A. {et~al.}, 2020, \mnras

\bibitem[{{Geringer-Sameth} {et~al}\mbox{.}(2015){Geringer-Sameth},
  {Koushiappas}, \& {Walker}}]{2015PhRvD..91h3535G}
{Geringer-Sameth} A., {Koushiappas} S.~M., {Walker} M.~G., 2015, \prd, 91,
  083535

\bibitem[{{Gregory} {et~al}\mbox{.}(2019){Gregory}, {Collins}, {Read}, {Irwin},
  {Ibata}, {Martin}, {McConnachie}, \& {Weisz}}]{2019MNRAS.485.2010G}
{Gregory} A.~L., {Collins} M. L.~M., {Read} J.~I., {Irwin} M.~J., {Ibata}
  R.~A., {Martin} N.~F., {McConnachie} A.~W., {Weisz} D.~R., 2019, \mnras, 485,
  2010

\bibitem[{{Hayashi} {et~al}\mbox{.}(2020){Hayashi}, {Chiba}, \&
  {Ishiyama}}]{2020arXiv200713780H}
{Hayashi} K., {Chiba} M., {Ishiyama} T., 2020, arXiv e-prints, arXiv:2007.13780

\bibitem[{{H{\'e}nault-Brunet} {et~al}\mbox{.}(2019){H{\'e}nault-Brunet},
  {Gieles}, {Sollima}, {Watkins}, {Zocchi}, {Claydon}, {Pancino}, \&
  {Baumgardt}}]{HenaultBrunet2019}
{H{\'e}nault-Brunet} V., {Gieles} M., {Sollima} A., {Watkins} L.~L., {Zocchi}
  A., {Claydon} I., {Pancino} E., {Baumgardt} H., 2019, \mnras, 483, 1400

\bibitem[{{Hubble}(1926)}]{Hubble26}
{Hubble} E.~P., 1926, \apj, 64, 321

\bibitem[{{Jeans}(1922)}]{1922MNRAS..82..122J}
{Jeans} J.~H., 1922, \mnras, 82, 122

\bibitem[{{Kamann} {et~al}\mbox{.}(2018){Kamann}, {Husser}, {Dreizler},
  {Emsellem}, {Weilbacher}, {Martens}, {Bacon}, {den Brok}, {Giesers},
  {Krajnovi{\'c}}, {Roth}, {Wendt}, \& {Wisotzki}}]{Kamann2018}
{Kamann} S. {et~al.}, 2018, \mnras, 473, 5591

\bibitem[{{Koposov} {et~al}\mbox{.}(2011){Koposov}, {Gilmore}, {Walker},
  {Belokurov}, {Evans}, {Fellhauer}, {Gieren}, {Geisler}, {Monaco}, {Norris},
  {Okamoto}, {Pe{\~n}arrubia}, {Wilkinson}, {Wyse}, \&
  {Zucker}}]{2011ApJ...736..146K}
{Koposov} S.~E. {et~al.}, 2011, \apj, 736, 146

\bibitem[{{Kowalczyk} {et~al}\mbox{.}(2013){Kowalczyk}, {{\L}okas},
  {Kazantzidis}, \& {Mayer}}]{2013MNRAS.431.2796K}
{Kowalczyk} K., {{\L}okas} E.~L., {Kazantzidis} S., {Mayer} L., 2013, \mnras,
  431, 2796

\bibitem[{{Lewis} \& {Bridle}(2002)}]{Lewis&Bridle02}
{Lewis} A., {Bridle} S., 2002, \prd, 66, 103511

\bibitem[{{Limousin} {et~al}\mbox{.}(2013){Limousin}, {Morand i}, {Sereno},
  {Meneghetti}, {Ettori}, {Bartelmann}, \& {Verdugo}}]{Limousin+13}
{Limousin} M., {Morand i} A., {Sereno} M., {Meneghetti} M., {Ettori} S.,
  {Bartelmann} M., {Verdugo} T., 2013, \ssr, 177, 155

\bibitem[{{Mamon} {et~al}\mbox{.}(2013){Mamon}, {Biviano}, \&
  {Bou{\'e}}}]{Mamon+13}
{Mamon} G.~A., {Biviano} A., {Bou{\'e}} G., 2013, \mnras, 429, 3079

\bibitem[{{Mamon} {et~al}\mbox{.}(2019){Mamon}, {Cava}, {Biviano}, {Moretti},
  {Poggianti}, \& {Bettoni}}]{2019A&A...631A.131M}
{Mamon} G.~A., {Cava} A., {Biviano} A., {Moretti} A., {Poggianti} B., {Bettoni}
  D., 2019, \aap, 631, A131

\bibitem[{{Mamon} \& {{\L}okas}(2005)}]{Mamon&Lokas05}
{Mamon} G.~A., {{\L}okas} E.~L., 2005, \mnras, 363, 705

\bibitem[{{Massari} {et~al}\mbox{.}(2018){Massari}, {Breddels}, {Helmi},
  {Posti}, {Brown}, \& {Tolstoy}}]{2018NatAs...2..156M}
{Massari} D., {Breddels} M.~A., {Helmi} A., {Posti} L., {Brown} A.~G.~A.,
  {Tolstoy} E., 2018, Nature Astronomy, 2, 156

\bibitem[{{Massari} {et~al}\mbox{.}(2020){Massari}, {Helmi}, {Mucciarelli},
  {Sales}, {Spina}, \& {Tolstoy}}]{2020A&A...633A..36M}
{Massari} D., {Helmi} A., {Mucciarelli} A., {Sales} L.~V., {Spina} L.,
  {Tolstoy} E., 2020, \aap, 633, A36

\bibitem[{{McConnachie} \& {C{\^o}t{\'e}}(2010)}]{2010ApJ...722L.209M}
{McConnachie} A.~W., {C{\^o}t{\'e}} P., 2010, \apjl, 722, L209

\bibitem[{{McMillan} \& {Binney}(2013)}]{2013MNRAS.433.1411M}
{McMillan} P.~J., {Binney} J.~J., 2013, \mnras, 433, 1411

\bibitem[{{Merrifield} \& {Kent}(1990)}]{1990AJ.....99.1548M}
{Merrifield} M.~R., {Kent} S.~M., 1990, \aj, 99, 1548

\bibitem[{{Merritt}(1985)}]{Merritt85}
{Merritt} D., 1985, \aj, 90, 1027

\bibitem[{{Merritt}(1987)}]{Merritt87}
{Merritt} D., 1987, \apj, 313, 121

\bibitem[{{Napolitano} {et~al}\mbox{.}(2014){Napolitano}, {Pota}, {Romanowsky},
  {Forbes}, {Brodie}, \& {Foster}}]{2014MNRAS.439..659N}
{Napolitano} N.~R., {Pota} V., {Romanowsky} A.~J., {Forbes} D.~A., {Brodie}
  J.~P., {Foster} C., 2014, \mnras, 439, 659

\bibitem[{{Noyola} {et~al}\mbox{.}(2008){Noyola}, {Gebhardt}, \&
  {Bergmann}}]{2008ApJ...676.1008N}
{Noyola} E., {Gebhardt} K., {Bergmann} M., 2008, \apj, 676, 1008

\bibitem[{{Osipkov}(1979)}]{1979PAZh....5...77O}
{Osipkov} L.~P., 1979, Pisma Astronomicheskii Zhurnal, 5, 77, english
  translation in Soviet Astron. Lett. 5, 42

\bibitem[{{Plummer}(1911)}]{1911MNRAS..71..460P}
{Plummer} H.~C., 1911, \mnras, 71, 460

\bibitem[{{Pontzen} {et~al}\mbox{.}(2013){Pontzen}, {Ro{\v s}kar}, {Stinson},
  \& {Woods}}]{2013ascl.soft05002P}
{Pontzen} A., {Ro{\v s}kar} R., {Stinson} G., {Woods} R., 2013, {pynbody:
  N-Body/SPH analysis for python}. Astrophysics Source Code Library

\bibitem[{{Posti} {et~al}\mbox{.}(2015){Posti}, {Binney}, {Nipoti}, \&
  {Ciotti}}]{2015MNRAS.447.3060P}
{Posti} L., {Binney} J., {Nipoti} C., {Ciotti} L., 2015, \mnras, 447, 3060

\bibitem[{{Read}(2014)}]{2014JPhG...41f3101R}
{Read} J.~I., 2014, Journal of Physics G Nuclear Physics, 41, 063101

\bibitem[{{Read} \& {Steger}(2017)}]{2017arXiv170104833R}
{Read} J.~I., {Steger} P., 2017, \mnras, 471, 4541

\bibitem[{{Read} {et~al}\mbox{.}(2018){Read}, {Walker}, \&
  {Steger}}]{2018MNRAS.481..860R}
{Read} J.~I., {Walker} M.~G., {Steger} P., 2018, \mnras, 481, 860

\bibitem[{{Read} {et~al}\mbox{.}(2019){Read}, {Walker}, \&
  {Steger}}]{2019MNRAS.484.1401R}
{Read} J.~I., {Walker} M.~G., {Steger} P., 2019, \mnras, 484, 1401

\bibitem[{{Read} {et~al}\mbox{.}(2006{\natexlab{a}}){Read}, {Wilkinson},
  {Evans}, {Gilmore}, \& {Kleyna}}]{2006MNRAS.tmp..153R}
{Read} J.~I., {Wilkinson} M.~I., {Evans} N.~W., {Gilmore} G., {Kleyna} J.~T.,
  2006{\natexlab{a}}, \mnras, 367, 387

\bibitem[{{Read} {et~al}\mbox{.}(2006{\natexlab{b}}){Read}, {Wilkinson},
  {Evans}, {Gilmore}, \& {Kleyna}}]{2006MNRAS.366..429R}
{Read} J.~I., {Wilkinson} M.~I., {Evans} N.~W., {Gilmore} G., {Kleyna} J.~T.,
  2006{\natexlab{b}}, \mnras, 366, 429

\bibitem[{{Richardson} \& {Fairbairn}(2014)}]{2014MNRAS.441.1584R}
{Richardson} T., {Fairbairn} M., 2014, \mnras, 441, 1584

\bibitem[{{Robotham} {et~al}\mbox{.}(2010){Robotham}, {Driver}, {Norberg},
  {Baldry}, {Bamford}, {Hopkins}, {Liske}, {Loveday}, {Peacock}, {Cameron},
  {Croom}, {Doyle}, {Frenk}, {Hill}, {Jones}, {van Kampen}, {Kelvin},
  {Kuijken}, {Nichol}, {Parkinson}, {Popescu}, {Prescott}, {Sharp}, {Sutherland
  }, {Thomas}, \& {Tuffs}}]{Robotham+10}
{Robotham} A. {et~al.}, 2010, PASA, 27, 76

\bibitem[{{Rojas-Ni{\~n}o} {et~al}\mbox{.}(2016){Rojas-Ni{\~n}o}, {Read},
  {Aguilar}, \& {Delorme}}]{2016MNRAS.459.3349R}
{Rojas-Ni{\~n}o} A., {Read} J.~I., {Aguilar} L., {Delorme} M., 2016, \mnras,
  459, 3349

\bibitem[{{Sanchis} {et~al}\mbox{.}(2004){Sanchis}, {{\L}okas}, \&
  {Mamon}}]{Sanchis+04}
{Sanchis} T., {{\L}okas} E.~L., {Mamon} G.~A., 2004, \mnras, 347, 1198

\bibitem[{{Sanders} \& {Evans}(2017)}]{Sanders&Evans17}
{Sanders} J.~L., {Evans} N.~W., 2017, \mnras, 472, 2670

\bibitem[{{Sanderson} {et~al}\mbox{.}(2017){Sanderson}, {Bellini}, {Casertano},
  {Lu}, {Melchior}, {Libralato}, {Bennett}, {Shao}, {Rhodes}, {Sohn},
  {Malhotra}, {Gaudi}, {Fall}, {Nelan}, {Guhathakurta}, {Anderson}, \&
  {Ho}}]{WFIRST2017}
{Sanderson} R.~E. {et~al.}, 2017, arXiv e-prints, arXiv:1712.05420

\bibitem[{{Spencer} {et~al}\mbox{.}(2018){Spencer}, {Mateo}, {Olszewski},
  {Walker}, {McConnachie}, \& {Kirby}}]{2018AJ....156..257S}
{Spencer} M.~E., {Mateo} M., {Olszewski} E.~W., {Walker} M.~G., {McConnachie}
  A.~W., {Kirby} E.~N., 2018, \aj, 156, 257

\bibitem[{{Strigari} {et~al}\mbox{.}(2007){Strigari}, {Bullock}, \&
  {Kaplinghat}}]{2007ApJ...657L...1S}
{Strigari} L.~E., {Bullock} J.~S., {Kaplinghat} M., 2007, \apjl, 657, L1

\bibitem[{{Syer} \& {Tremaine}(1996)}]{1996MNRAS.282..223S}
{Syer} D., {Tremaine} S., 1996, \mnras, 282, 223

\bibitem[{{Theia Collaboration} {et~al}\mbox{.}(2017){Theia Collaboration},
  {Boehm}, {Krone-Martins}, {Amorim}, {Anglada-Escude}, {Brandeker}, {Courbin},
  {Ensslin}, {Falcao}, {Freese}, {Holl}, {Labadie}, {Leger}, {Malbet}, {Mamon},
  {McArthur}, {Mora}, {Shao}, {Sozzetti}, {Spolyar}, {Villaver}, {Albertus},
  {Bertone}, {Bouy}, {Boylan-Kolchin}, {Brown}, {Brown}, {Cardoso}, {Chemin},
  {Claudi}, {Correia}, {Crosta}, {Crouzier}, {Cyr-Racine}, {Damasso}, {da
  Silva}, {Davies}, {Das}, {Dayal}, {de Val-Borro}, {Diaferio}, {Erickcek},
  {Fairbairn}, {Fortin}, {Fridlund}, {Garcia}, {Gnedin}, {Goobar}, {Gordo},
  {Goullioud}, {Hambly}, {Hara}, {Hobbs}, {Hog}, {Holland}, {Ibata}, {Jordi},
  {Klioner}, {Kopeikin}, {Lacroix}, {Laskar}, {Le Poncin-Lafitte}, {Luri},
  {Majumdar}, {Makarov}, {Massey}, {Mennesson}, {Michalik}, {Moitinho de
  Almeida}, {Mourao}, {Moustakas}, {Murray}, {Muterspaugh}, {Oertel},
  {Ostorero}, {Perez-Garcia}, {Platais}, {de Mora}, {Quirrenbach}, {Randall},
  {Read}, {Regos}, {Rory}, {Rybicki}, {Scott}, {Schneider}, {Scholtz},
  {Siebert}, {Tereno}, {Tomsick}, {Traub}, {Valluri}, {Walker}, {Walton},
  {Watkins}, {White}, {Evans}, {Wyrzykowski}, \& {Wyse}}]{2017arXiv170701348T}
{Theia Collaboration} {et~al.}, 2017, arXiv e-prints, arXiv:1707.01348

\bibitem[{{Ural} {et~al}\mbox{.}(2015){Ural}, {Wilkinson}, {Read}, \&
  {Walker}}]{2015NatCo...6E7599U}
{Ural} U., {Wilkinson} M.~I., {Read} J.~I., {Walker} M.~G., 2015, Nature
  Communications, 6, 7599

\bibitem[{{van der Marel}(1994)}]{vanderMarel94}
{van der Marel} R.~P., 1994, \mnras, 270, 271

\bibitem[{{van der Marel} \& {Anderson}(2010)}]{2010ApJ...710.1063V}
{van der Marel} R.~P., {Anderson} J., 2010, \apj, 710, 1063

\bibitem[{{Vasiliev}(2019)}]{Vasiliev19}
{Vasiliev} E., 2019, \mnras, 482, 1525

\bibitem[{{Verolme} {et~al}\mbox{.}(2002){Verolme}, {Cappellari}, {Copin}, {van
  der Marel}, {Bacon}, {Bureau}, {Davies}, {Miller}, \& {de
  Zeeuw}}]{2002MNRAS.335..517V}
{Verolme} E.~K. {et~al.}, 2002, \mnras, 335, 517

\bibitem[{{Vitral} \& {Mamon}(2020)}]{Vitral&Mamon20}
{Vitral} E., {Mamon} G.~A., 2020, arXiv e-prints, arXiv:2010.05532

\bibitem[{{Walker} {et~al}\mbox{.}(2009){Walker}, {Mateo}, \&
  {Olszewski}}]{2009AJ....137.3100W}
{Walker} M.~G., {Mateo} M., {Olszewski} E.~W., 2009, \aj, 137, 3100

\bibitem[{{Walker} \& {Pe{\~n}arrubia}(2011)}]{2011ApJ...742...20W}
{Walker} M.~G., {Pe{\~n}arrubia} J., 2011, \apj, 742, 20

\bibitem[{{Watkins} {et~al}\mbox{.}(2013){Watkins}, {van de Ven}, {den Brok},
  \& {van den Bosch}}]{Watkins2013}
{Watkins} L.~L., {van de Ven} G., {den Brok} M., {van den Bosch} R. C.~E.,
  2013, \mnras, 436, 2598

\bibitem[{{Watkins} {et~al}\mbox{.}(2015{\natexlab{a}}){Watkins}, {van der
  Marel}, {Bellini}, \& {Anderson}}]{Watkins2015b}
{Watkins} L.~L., {van der Marel} R.~P., {Bellini} A., {Anderson} J.,
  2015{\natexlab{a}}, \apj, 812, 149

\bibitem[{{Watkins} {et~al}\mbox{.}(2015{\natexlab{b}}){Watkins}, {van der
  Marel}, {Bellini}, \& {Anderson}}]{Watkins2015a}
{Watkins} L.~L., {van der Marel} R.~P., {Bellini} A., {Anderson} J.,
  2015{\natexlab{b}}, \apj, 803, 29

\bibitem[{{White} \& {Shawl}(1987)}]{White&Shawl87}
{White} R.~E., {Shawl} S.~J., 1987, \apj, 317, 246

\bibitem[{{Wojtak} {et~al}\mbox{.}(2009){Wojtak}, {{\L}okas}, {Mamon}, \&
  {Gottl{\"o}ber}}]{2009MNRAS.399..812W}
{Wojtak} R., {{\L}okas} E.~L., {Mamon} G.~A., {Gottl{\"o}ber} S., 2009, \mnras,
  399, 812

\bibitem[{{York} {et~al}\mbox{.}(2000){York}, {Adelman}, {Anderson},
  {Anderson}, {Annis}, {Bahcall}, {Bakken}, {Barkhouser}, {Bastian}, {Berman},
  {Boroski}, {Bracker}, {Briegel}, {Briggs}, {Brinkmann}, {Brunner}, {Burles},
  {Carey}, {Carr}, {Castand er}, {Chen}, {Colestock}, {Connolly}, {Crocker},
  {Csabai}, {Czarapata}, {Davis}, {Doi}, {Dombeck}, {Eisenstein}, {Ellman},
  {Elms}, {Evans}, {Fan}, {Federwitz}, {Fiscelli}, {Friedman}, {Frieman},
  {Fukugita}, {Gillespie}, {Gunn}, {Gurbani}, {de Haas}, {Haldeman}, {Harris},
  {Hayes}, {Heckman}, {Hennessy}, {Hindsley}, {Holm}, {Holmgren}, {Huang},
  {Hull}, {Husby}, {Ichikawa}, {Ichikawa}, {Ivezi{\'c}}, {Kent}, {Kim},
  {Kinney}, {Klaene}, {Kleinman}, {Kleinman}, {Knapp}, {Korienek}, {Kron},
  {Kunszt}, {Lamb}, {Lee}, {Leger}, {Limmongkol}, {Lindenmeyer}, {Long},
  {Loomis}, {Loveday}, {Lucinio}, {Lupton}, {MacKinnon}, {Mannery}, {Mantsch},
  {Margon}, {McGehee}, {McKay}, {Meiksin}, {Merelli}, {Monet}, {Munn},
  {Narayanan}, {Nash}, {Neilsen}, {Neswold}, {Newberg}, {Nichol}, {Nicinski},
  {Nonino}, {Okada}, {Okamura}, {Ostriker}, {Owen}, {Pauls}, {Peoples},
  {Peterson}, {Petravick}, {Pier}, {Pope}, {Pordes}, {Prosapio},
  {Rechenmacher}, {Quinn}, {Richards}, {Richmond}, {Rivetta}, {Rockosi},
  {Ruthmansdorfer}, {Sand ford}, {Schlegel}, {Schneider}, {Sekiguchi},
  {Sergey}, {Shimasaku}, {Siegmund}, {Smee}, {Smith}, {Snedden}, {Stone},
  {Stoughton}, {Strauss}, {Stubbs}, {SubbaRao}, {Szalay}, {Szapudi}, {Szokoly},
  {Thakar}, {Tremonti}, {Tucker}, {Uomoto}, {Vand en Berk}, {Vogeley},
  {Waddell}, {Wang}, {Watanabe}, {Weinberg}, {Yanny}, {Yasuda}, \& {SDSS
  Collaboration}}]{York+00}
{York} D.~G. {et~al.}, 2000, \aj, 120, 1579

\bibitem[{{Zhao}(1996)}]{1996MNRAS.278..488Z}
{Zhao} H., 1996, \mnras, 278, 488

\bibitem[{{Zhu} {et~al}\mbox{.}(2016){Zhu}, {van de Ven}, {Watkins}, \&
  {Posti}}]{2016MNRAS.463.1117Z}
{Zhu} L., {van de Ven} G., {Watkins} L.~L., {Posti} L., 2016, \mnras, 463, 1117

\bibitem[{{Zocchi} {et~al}\mbox{.}(2017){Zocchi}, {Gieles}, \&
  {H{\'e}nault-Brunet}}]{2017MNRAS.468.4429Z}
{Zocchi} A., {Gieles} M., {H{\'e}nault-Brunet} V., 2017, \mnras, 468, 4429

\end{thebibliography}

\end{document}